\documentclass[
   10pt,sigconf,letterpaper
]{acmart}
\usepackage{booktabs,tabularx}

\usepackage[mode=text, math-rm=\text, detect-all, binary-units=true, per-mode=symbol, range-phrase=\,--\,, range-units=single, detect-weight=true, detect-family=true]{siunitx}
\DeclareSIUnit{\bit}{bit}
\DeclareSIUnit{\bps}{\bit\per\s}

\usepackage{tikz}
\usepackage{pgfplots}
\pgfplotsset{compat=1.16}
\usetikzlibrary{calc,arrows.meta}
\usepackage{subcaption}
\usepackage{dblfloatfix} 

\usepackage{csquotes}

\graphicspath{{figures}}

\usepackage[nice]{nicefrac} % Display ½ nicely
\usepackage{xcolor}         % Colors!
\usepackage{xifthen}
\usepackage{mathtools}
\usepackage{multicol}
\usepackage{placeins}

\usepackage[capitalize]{cleveref}

\crefname{section}{\S}{Sections}
\Crefname{section}{Section}{Sections}
\crefformat{section}{\S#2#1#3}
\Crefformat{section}{Section~#2#1#3}
\crefformat{footnote}{#2\footnotemark[#1]#3}

\usepackage{enumitem}       % Allows customizing enumerate counters
\newcommand{\enumfont}{\bfseries\sffamily}
\SetEnumitemKey{myenum}{
}
\newlist{benenum}{enumerate}{1}
\setlist[benenum]{myenum, label={\enumfont B\arabic*}, ref=B\arabic*}
\crefname{benenumi}{benefit}{benefits}

\newtheorem{insight}{Insight}

\usepackage[normalem]{ulem}
\newcommand{\gennote}[5][blue]{{\color{#1}%
		$\rule{8pt}{8pt}_\textsf{\bfseries #2}^\textsf{\bfseries #3}$
		\textcolor{gray}{\emph{\sout{#4}}}#5}%
}

\newcommand{\authorcomment}[3]{%
	\expandafter\newcommand\csname#1\endcsname[1]{\gennote[#3]{#2}{}{}{##1}}%
	\expandafter\newcommand\csname#1S\endcsname[2][]{\gennote[#3]{#2}{}{##1}{##2}}%
	\expandafter\newcommand\csname#1Q\endcsname[1]{\gennote[#3]{#2}{Question}{}{##1}}%
	\expandafter\newcommand\csname#1N\endcsname[1]{\gennote[#3]{#2}{Note}{}{##1}}%
	\expandafter\newcommand\csname#1C\endcsname[1]{\gennote[#3]{#2}{Comment}{}{##1}}%
	\expandafter\newcommand\csname#1T\endcsname[1]{\gennote[#3]{#2}{TODO}{}{##1}}%
}
\authorcomment{stefan}{StS}{blue}
\authorcomment{simon}{SiS}{orange}
\authorcomment{adrian}{AP}{red}
\authorcomment{ml}{ML}{purple}

\newcommand{\funarg}[1]{\ifthenelse{\isempty{#1}}{}{\!\left(#1\right)}}

\newcommand{\link}{\ensuremath{\ell}}
\newcommand{\lcap}[1][\link]{\ensuremath{C_{#1}}}
\newcommand{\lbuf}[1][\link]{\ensuremath{B_{#1}}}
\newcommand{\propd}[1][\link]{\ensuremath{d_{#1}}}
\newcommand{\propdf}[1][i,\link]{\ensuremath{d_{#1}^{\smash{\mathrm{f}}}}}
\newcommand{\propdb}[1][i,\link]{\ensuremath{d_{#1}^{\smash{\mathrm{b}}}}}
\newcommand{\propdp}[1][i]{\ensuremath{d_{#1}^{\smash{\mathrm{p}}}}}
\newcommand{\pathi}[1][i]{\ensuremath{\pi_{#1}}}
\newcommand{\users}[1][\link]{\ensuremath{U_{#1}}}

\newcommand{\cwnd}[2][i]{\ensuremath{w_{#1}\funarg{#2}}}
\newcommand{\lat}[2][\link]{\ensuremath{\tau_{#1}\funarg{#2}}}
\newcommand{\latpath}[2][i]{\ensuremath{\tau_{\pathi[#1]}\funarg{#2}}}
\newcommand{\rate}[2][i]{\ensuremath{x_{#1}\funarg{#2}}}

\newcommand{\arriv}[2][\link]{\ensuremath{y_{#1}\funarg{#2}}}
\newcommand{\queue}[2][\link]{\ensuremath{q_{#1}\funarg{#2}}}
\newcommand{\loss}[2][\link]{\ensuremath{p_{#1}\funarg{#2}}}

\newcommand{\dqueue}[2][\link]{\ensuremath{\dot q_{#1}\funarg{#2}}}

\newtheorem{theorem}{Theorem}
\newtheorem{definition}{Definition}

\acmConference[IMC '22]{IMC '22: ACM Internet Measurement Conference}{October 25--27, 2022}{Nice, France}
\setcopyright{none}
\copyrightyear{}
\acmYear{}
\acmDOI{}
\acmPrice{}
\acmISBN{}
\renewcommand\footnotetextcopyrightpermission[1]{} 

\begin{document}

\title[Performance, Fairness, and Stability of BBR]{Model-Based Insights on the\\Performance, Fairness, and Stability of BBR}

\author{Simon Scherrer}
\email{simon.scherrer@inf.ethz.ch}
\affiliation{%
  \institution{ETH Zurich}
  \country{Switzerland}
}
\author{Markus Legner}
\email{markus.legner@inf.ethz.ch}
\affiliation{%
  \institution{ETH Zurich}
  \country{Switzerland}
}
\author{Adrian Perrig}
\email{adrian.perrig@inf.ethz.ch}
\affiliation{%
  \institution{ETH Zurich}
  \country{Switzerland}
}
\author{Stefan Schmid}
\email{stefan_schmid@univie.ac.at}
\affiliation{%
  \institution{TU Berlin/Uni. Vienna}
  \country{Germany/Austria}
}

% \author{Anonymous}

% \renewcommand{\shortauthors}{}

\begin{abstract}
    % The explosive growth of network traffic and
    % the increasing diversity of applications and paths
    % in the Internet has recently reignited the quest for efficient,
    % fair, and flexible congestion-control algorithms (CCAs).
    % However, the diverse network settings found in the Internet
    %  pose a persistent challenge for the
    % comprehensive experimental evaluation of new CCAs; for example, 
    % undesirable properties of BBR, a new CCA, were only discovered
    % after it had already been deployed.
    % By complementing experiment-based designs
    % and evaluations of new CCAs, analytical models
    % can help meet the stringent 
    % fairness and performance requirements 
    % for CCAs.
    
    % This paper pursues such an analytical approach
    % to evaluate CCAs using fluid models (based on differential equations),
    % which can be constructed and adapted flexibly and
    % hence provide fast insights already during the development
    % of a new CCA. In particular, we present fluid models for BBRv1 and BBRv2,
    % which do not naturally fit into the existing fluid frameworks
    % because these CCAs estimate the bottleneck bandwidth through a periodic
    % sequence of probing phases.
    % Our approach allows us to perform extensive
    % systematic simulations of BBR by itself or 
    % coexisting with traditional CCAs under diverse conditions.
    % Our experimentally validated fluid model
    % provides several new insights into
    % fundamental metrics of both BBR versions,
    % including fairness and efficiency.
    
    Google's BBR is the most prominent result
    of the recently revived quest for efficient, fair, and
    flexible congestion-control algorithms (CCAs).
    While the performance of BBR has been investigated
    by numerous studies, previous work still leaves gaps
    in the understanding of BBR performance: Experiment-based
    studies generally only consider network settings that
    researchers can set up with manageable effort, and
    model-based studies neglect important issues like convergence.
    
    To complement previous BBR analyses, this paper presents
    a fluid model of BBRv1 and BBRv2, allowing both efficient
    simulation under a wide variety of network settings and
    analytical treatment such as stability analysis.
    By experimental validation, we show that our fluid model
    provides highly accurate predictions of BBR behavior.
    Through extensive simulations and theoretical analysis,
    we arrive at several insights into both BBR versions, 
    including a previously unknown bufferbloat issue in BBRv2.
    \vspace{-20pt}
\end{abstract}

% \keywords{congestion control, fluid models, modeling, BBR, fairness}

\maketitle

%%%%%%%%%%%%%%%%%%%%%%
\section{Introduction}
\label{sec:intro}

\begin{figure}
    \centering
    \includegraphics[width=\linewidth]{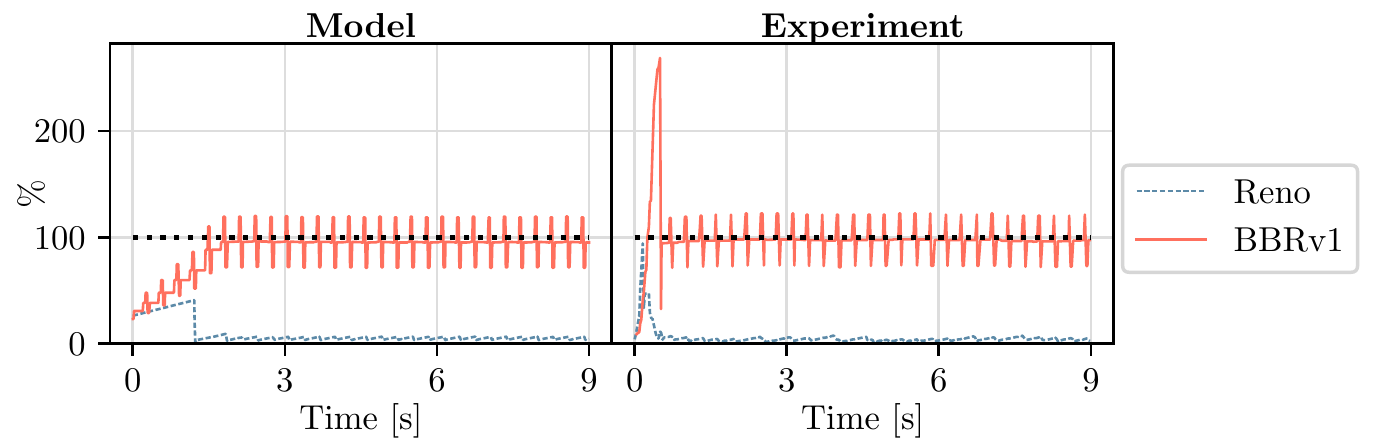}
    \vspace{-20pt}
    \caption{Competition of sending rates (in \% of link bandwidth)
    between a Reno flow and a BBRv1 flow,
    according to our fluid model and experiment data. 
    % The fluid model abstracts away algorithm features (e.g.,
    % startup phase), but is highly predictive
    % of the basic CCA property (here: unfairness of BBRv1
    % towards Reno).
    }
    \label{fig:introduction:fluid-model-visualization}
    \vspace{-20pt}
\end{figure}

To this day, ever changing applications, traffic patterns, network capacities, and path types 
prompt research into new and better congestion-control algorithms (CCAs).
Most prominent in recent years was Google's introduction of BBR~\cite{cardwell2016bbr}, which was promptly enabled in 2017 for some Google services and thus widely deployed in the public Internet~\cite{cardwell2017bbr-ietf}.
Since then, several theoretical and experimental studies of the behavior of BBR~\cite{ware2019modeling,hock2017experimental,dong2018pcc,scholz2018towards,turkovic2019fifty} have identified issues with 
this first version of BBR, 
relating to both fairness (especially towards loss-based CCAs) and efficiency (e.g., excessive queue buildup).
As a result, BBRv2~\cite{cardwell2019bbrv2} has been proposed, triggering
another series of evaluation studies~\cite{gomez2020performance,kfoury2020emulation,nandagiri2020bbrvl,song2021understanding}.
% Unfortunately, most studies of this new CCA were made only \emph{after} it had already been deployed for a large portion of traffic in the public Internet.

Still, the characterization of BBR performance remains incomplete. 
Experiment-based studies~\cite{hock2017experimental,turkovic2019fifty,scholz2018towards,kfoury2020emulation,nandagiri2020bbrvl,gomez2020performance},
by their nature, allow statements relating to the concrete
network settings in the experiments. Given the variety of scenarios
in which a CCA might be deployed, such experimental investigations
could only be made exhaustive with great effort and 
large-scale testbeds which only a minority of researchers has access to.
Previous model-based studies contain BBR steady-state models 
that are highly valuable for specific settings (e.g., deep buffers~\cite{ware2019beyond}
or wireless links~\cite{yang2019adaptive}); however, 
a deep theoretical understanding of BBR also requires a model
that is valid for general settings and allows investigation
of the convergence process.

In this paper, we complement previous approaches
to BBR analysis with a classic approach in CCA research:
fluid models consisting of differential equations~\cite{low2002internet,vardoyan2018towards,srikant2004mathematics,liu2003fluid,misra2000fluid,foreest2003analysis,raina2005buffer}.
Such fluid models are unique in their suitability for \emph{both} efficient simulation
and theoretical stability analysis. Enabling efficient simulation is critical
because the model must be simulated under a plethora of configurations,
including settings that are expensive to build. 
Enabling theoretical stability analysis is
crucial because the equilibria (i.e., steady states) of the CCA dynamics 
are only relevant for performance characterization if stable in a
control-theoretic sense, i.e.,  if the dynamics actually converge to the equilibria.

While a fluid model for BBR is thus well-suited to complement previous work,
constructing such a model is challenging because BBR does not naturally fit
into the existing fluid-model framework for loss-based CCAs~\cite{low2002internet,vardoyan2018towards}.
In fact, BBR does not exclusively rely on a congestion window affected by loss, 
but includes traffic pulses for capacity probing
and measurement-driven state transitions.
By using new techniques, e.g., by mimicking the probing pulses with sigmoid functions,
this work establishes the first highly accurate and highly general model of BBR,
both for versions 1 and 2.

Our fluid model predicts BBR behavior with
high accuracy, which we validate with experiments by means of the network emulator mininet~\cite{lantz2010network}.
The validated model allows us to perform extensive systematic simulations of BBR by itself or coexisting with traditional CCAs,
confirming BBR perforrmance issues from previous studies and yielding new insights.
Moreover, we apply dynamical-system analysis (i.e., the Lyapunov method)
to our fluid model to identify asymptotically stable equilibria of the BBR dynamics.

Our main contributions are the following:
\begin{itemize}[topsep=5pt]
    \item We introduce the first accurate and general fluid model for BBR (versions 1 and 2),
    using new techniques such as sigmoid pulses and mode variables.
    \item We present extensive, systematic model-based calculations, experimentally validate the results of these calculations, and provide profound insights into fundamental metrics of BBR.
    \item We analytically identify asymptotically stable equilibria of BBRv1 and BBRv2.
    \item We confirm both experimentally and analytically that BBRv1 can lead to unfair bandwidth allocations---espe-cially when it competes against loss-based CCAs, but also in competition with itself in certain settings.
    \item We confirm that BBRv2 eliminates most of the undesirable behavior of BBRv1, but we also
    identify settings in which BBRv2 leads to bufferbloat and unfairness.
%     \item We present insights into strengths and limitations of fluid
% models for modeling the performance
% of CCAs.
    % \item To ensure reproducibility and facilitate follow-up work on additional analysis, 
    % we will make our fluid-model simulation implementation 
    % publicly available.
\end{itemize}

% \input{sections/motivation}
%%%%%%%%%%%%%%%%%%%%%%
\section{Network Fluid Model}
\label{sec:models}

In this section, we present our network fluid model, which closely follows 
the work by Low et al.~\cite{low2002internet}. 
However, we have made several improvements to the network model, which we will highlight in the following.
We denote function~$f(t)$ by $f$ and its derivative by~$\dot{f}$,
unless the argument of the function differs
from the default time variable~$t$.

In our model, the network consists of links \link{} with 
capacity \lcap{}, buffer size \lbuf{}, and 
propagation delay~\propd{}.

\paragraph{Link-Arrival Rate} 
The arrival rate~\arriv{t} at link~\link{} is
\begin{equation}
    \arriv{} = \sum_{i \in \users} \rate{t-\propdf},
    \label{eq:arriv}
\end{equation}
where \users{} is the set of agents using link~\link{}, \rate{t} is the sending
rate of agent~$i$ at time~$t$, 
and~$\propdf{}$ is the propagation delay 
between agent~$i$ and link~\link{}.
Together with Low et al.~\cite{low2002internet},
we neglect queuing delay and packet losses previous to
link~$\ell$.
% Note that using the propagation delay~\propdf{} as feedback delay is a
% simplification; in reality, the feedback delay also contains queuing delay
% and would be more accurately captured by~\latpath[i,\link]{t'}, where~\pathi[i,\ell]
% is the path from agent~$i$ up to link~\link{} and~$t'=t-\latpath[i,\link]{t'}$ is the time when
% agent~$i$ sent the traffic reaching link~\link{} at time~$t$.
% As this circular dependency is analytically difficult,
% Low et al.'s simplification suggests a fixed quantity as feedback delay). 
% Moreover, the arrival rate~\arriv{t} technically should take
% into account packet drops on links previous to~\link{}.
% Together with Low et al., however,
% we ignore such effects on the arrival rate and model
% loss dynamics differently (cf.~\cref{sec:de-system:loss}).

\paragraph{Queue Length}
\label{sec:de-system:queue}
In general, the queue length grows or shrinks
according to the discrepancy
between combined arrival rate~$y_\ell$ and the
transmission capacity~$\lcap{}$ at the respective
link~\cite{low2002internet}, but never exceeds
its buffer size~$\lbuf{}$:
\begin{equation}
  \dqueue{} = \left(1-\loss{}\right) \cdot \arriv{} - \lcap{}, \quad \queue{t} \in [0, \lbuf],
\end{equation}
where~\loss{t} is the loss probability
of link~\link{} at time~$t$ (cf.~\cref{sec:de-system:loss}).
We have refined the model by Low et al.\
to additionally capture the
effect of packet drops on the queue length.

% The resulting queue dynamics are
% \begin{equation}
%     \dqueue{t} =
%     \begin{cases}
%       \min\{\Delta(t),0\} & \text{if } \queue{t} = \lbuf{}\\
%       \max\{\Delta(t),0\} & \text{if } \queue{t} = 0\\
%       \Delta(t) & \text{otherwise},
%     \end{cases}
%     \label{eq:ql}
% \end{equation}
% which ensures that $\queue{t} \in [0, \lbuf]$ at all times.

\paragraph{Latency}
The link latency is the fixed link propagation delay
plus the queuing delay, which depends on
queue size~$\queue{t}$.
The latency of a path is
the sum of link latencies:
\begin{align}
    \latpath{} = \sum_{\link \in \pi_i} \lat{} = 
    \sum_{\link \in \pi_i} d_\link + \frac{\queue{}}{\lcap}.
    \label{eq:taul}
\end{align}

\paragraph{Loss Probability and Queuing Disciplines}
\label{sec:de-system:loss}
Without active queuing discipline, loss occurs if the buffer of
a link is full. Given such a simple \emph{drop-tail} policy,
the loss probability is given by the relative excess rate
whenever the queue is full, and is 0
otherwise~\cite{wischik2006queueing}.
To facilitate analytical treatment, we refine previous models by a smooth approximation:
% \begin{equation}
%     \loss{t} = \begin{cases}1 - \frac{\lcap{}}{\arriv{t}} & \text{if } \queue{t} = \lbuf{} \text{ and } y_{\link} > \lcap{}\\0 & \text{otherwise}\end{cases} \hspace{30pt} \loss{t} \in [0, 1]
%     \label{eq:pl:taildrop}
% \end{equation}
\begin{equation}
    \loss{t} = 
        \sigma\big(\arriv{t}-\lcap\big) \cdot \left(1 - \frac{\lcap{}}{\arriv{}}\right)\cdot\left(\frac{\queue{}}{\lbuf{}}\right)^L
    \label{eq:pl:taildrop}
\end{equation}
where~$L \gg 1$ and~$\sigma(v)$ is
a relatively sharp sigmoid function:
\begin{equation}
    \sigma(v) = \frac{1}{1+\mathrm{e}^{-K\cdot v}}
    \label{eq:sigmoid}
\end{equation} with~$K \gg 1$ controlling
the sharpness of the increase at~$v = 0$.
% Note that $\queue{t} \in [0, \lbuf]$ implies  $\loss{} \in [0, 1]$.

In contrast to drop-tail, the loss probability
under the RED queuing discipline
moves synchronously with the queue size.
% according to configuration parameters as follows:
% \begin{equation}
%     \loss{t} =
%     \begin{cases}
%         0 & \text{if } \queue{t} \leq \underline{q}_\link\\
%         \frac{\overline{p}_\link}{\overline{q}_\link - \underline{q}_\link} \cdot \left(\queue{t} - \underline{q}_\link(t)\right) & \text{if } \queue{t} \in (\underline{q}_\link, \overline{q}_\link)\\
%         \frac{1}{\overline{q}_\link} \cdot \left(\queue{t} - \overline{q}_\link\right) + \frac{\overline{p}_\link}{\overline{q}_\link}\left(2\overline{q}_\link - \queue{t}\right) & \text{if } \queue{t} \in [\overline{q}_\link, 2\overline{q}_\link)\\
%         1 & \text{if } \queue{t} \geq 2\overline{q}_\link,
%     \end{cases}
%     \label{eq:pl:red:full}
% \end{equation} where~$\overline{p}_\link$, $\overline{q}_\link$
% and~$\underline{q}_\link$ are configurable parameters.
% For the sake of simplicity, 
We approximate RED behavior
as follows:
% which corresponds to~\cref{eq:pl:red:full}
% with~$\overline{p}_\link = 1$, $\overline{q}_\link = \lbuf{}$,
% and~$\underline{q}_\link = 0$:
\begin{equation}
    \loss{} = \frac{\queue{}}{\lbuf{}} \quad\in[0, 1].
    \label{eq:pl:red}
\end{equation}

Regarding the loss probability of paths,
link-specific loss probabilities are assumed to be
small enough such that the following approximations
regarding loss hold:
\begin{equation}
    p_{\pathi}(t) = 1 - \prod_{\link \in \pathi} \left(1 - \loss{t+\propdf}\right) \approx \sum_{\link\in\pathi} \loss{t+\propdf}.
    \label{eq:ppi}
\end{equation}

\paragraph{Congestion Window and Sending Rate}
\label{model-cwnd-rate}
For the window-based congestion-control
algorithms Reno and CUBIC (cf.~\cref{sec:model-cc}),
the sending rate of agent~$i$ is determined by the congestion-window size~$w_i$ and round-trip latency:
\begin{equation}
    \rate{} = \frac{\cwnd{}}{\lat[i]{}}.
\end{equation}

% In contrast, BBR is not only based on a congestion
% window. We will model the sending rate adaptations
% of BBR in \cref{sec:model-bbr}.

\section{BBR Fluid Model}
\label{sec:model-bbr}

In this section, we introduce the first fluid model for
BBR, both for BBRv1~\cite{cardwell2016bbr} and
BBRv2~\cite{cardwell2019bbrv2}.
Interestingly, the fluid-model techniques
used for the  loss-based CCAs 
cannot reflect essential
BBR features, in particular its phases with different
behavior. Hence, we construct
our BBR model using new
techniques, i.e.,
periodic probing pulses and mode variables
(for simulating the BBR state machine).
In the following, we first describe
the behavior of BBR for
both versions 1 and 2. Then, we
present our fluid model for BBR
by means of a basic fluid model,
which can be concretized for
each version.

\subsection{Description of BBR}
\label{sec:model:bbr:description}
Fundamentally, BBR continuously performs
measurements to estimate two core properties
of the network path, namely the bottleneck
bandwidth~\texttt{BtlBw} and 
the minimal round-trip time (RTT)~\texttt{RTprop}
(as given by the propagation delay).
To estimate these properties, BBR constantly
switches between two states, 
namely the \texttt{ProbeBW} state
and the \texttt{ProbeRTT} state.
While the \texttt{ProbeBW} state consumes
most of flow lifetime and is considerably
different across the two BBR versions,
the \texttt{ProbeRTT} state is only
infrequently and briefly entered
and is mostly identical
across both BBR versions.

\textbf{\texttt{ProbeRTT} state.} BBR enters the \texttt{ProbeRTT}
state if no smaller round-trip time
than the existing \texttt{RTprop} estimate
is observed for 10 seconds. 
To discover the propagation delay, 
the \texttt{ProbeRTT} state tries to eliminate 
queuing delay by
restricting the data in flight 
(\emph{inflight} in BBR terminology)
to a small volume during 200 ms. 
In BBRv1, this small
volume has a fixed size of 4 segments;
since this volume has been found
to be too conservative,
the \texttt{ProbeRTT} inflight limit
in BBRv2 has been chosen to half the
estimated bandwidth-delay product,
i.e., half the product of \texttt{BtlBw}
and \texttt{RTprop}.

\textbf{\texttt{ProbeBW} state in BBRv1.} 
The \texttt{ProbeBW} state aims
at measuring the bottleneck bandwidth
of the network path, and includes a
periodic probing strategy
with the \emph{pacing rate}
as the primary control of 
the sending rate.
In this probing strategy,
each period consists of 8 phases with the
duration of \texttt{RTprop}. 
In one phase randomly chosen from
the first 7 phases, BBRv1
sets its pacing rate to $\nicefrac54\cdot\texttt{BtlBw}$
to find the capacity limit of
the path.
In the subsequent phase, BBRv1
decreases its pacing rate
to $\nicefrac34\cdot\texttt{BtlBw}$
to drain the queues
potentially built up during the
aggressive previous phase.
In the other 6 phases of the period,
BBRv1 paces at rate \texttt{BtlBw}.
At the period end,
the maximum delivery rate 
from the period is
then considered the new
bottleneck bandwidth estimate
and thus serves as a base
pacing rate for the next
period.

\textbf{BBRv1 congestion window.} While the focus on the pacing rate
distinguishes BBRv1 from window-based
CCAs like Reno or CUBIC,
BBRv1 accidentally degenerates
into a window-based CCA
in certain circumstances. In particular,
BBRv1 also maintains a congestion
window, which amounts to twice
the estimated BDP and was intended
as a safeguard against
`common network pathologies'
such as delayed
ACKs~\cite{cardwell2016bbr}.
Contrary to design intention,
this inflight limit by the congestion window 
is  the essential constraint
on the sending rate of BBRv1
when competing with loss-based
CCAs given large 
buffers~\cite{hock2017experimental,ware2019modeling}.

\textbf{\texttt{ProbeBW} state in BBRv2.} This unintentional relevance
of the inflight limit
in some circumstances, plus
the unfairness
towards loss-based CCAs in
other situations, led
Google to revise the
\texttt{ProbeBW} mechanism for BBRv2.
This revision mainly aimed at
making BBR less aggressive,
through increasing its
sensitivity to loss and
ECN signals (where we henceforth
only consider loss for simplicity), less frequent
probing, and a persistent coupling
between inflight limits and the
sending rate. To be precise,
BBRv2 tries to obtain additional
bandwidth only every few seconds,
where the time between such
probings is given by the minimum
of 62 estimated RTTs (chosen
for fairness reasons) and
a random value between 2 and 3
seconds. In this probing,
BBRv2 first paces at the rate given
by \texttt{BtlBw} for one \texttt{RTprop},
with the goal to achieve an inflight
corresponding to the bandwidth-delay product. 
Then, BBRv2 sets its pacing rate to
$\nicefrac54\cdot\texttt{BtlBw}$ and
increases the inflight until it
reaches $\nicefrac54$ of the estimated
BDP \emph{or} the loss rate exceeds~2\%.
At this point, the bottleneck-bandwidth
estimate \texttt{BtlBw} is updated
to the maximum delivery
rate from the last two \texttt{ProbeBW}
periods. Moreover, BBRv2 also records
the maximum tenable inflight
in state variable~\texttt{inflight\_hi},
which tracks the observed inflight,
but is reduced by a multiplicative decrease~$\beta$ 
if the exponential-increase phase
has been terminated by excessive loss.
Afterwards, BBRv2 chooses a pacing
rate of~$\nicefrac34\cdot\texttt{BtlBw}$
until the inflight is reduced 
to an arguably safe level,
which corresponds to the minimum of
the estimated BDP and 85\% of 
the previously measured \texttt{inflight\_hi} 
(where the erased 15\% are termed
\texttt{headroom} in BBRv2).
Once the inflight has been reduced
to that level, BBRv2 enters into
\emph{cruising} mode.
In cruising mode, BBRv2
aims to keep its inflight on a safe level
by introducing the additional inflight
bound \texttt{inflight\_lo}, which is
activated if packet loss occurs:
\texttt{inflight\_lo} starts 
from the congestion-window size at the moment
of loss and is reduced by~$\beta$ upon packet loss.
In contrast to \texttt{inflight\_hi},
which serves as a \emph{long-term} 
inflight bound,~\texttt{inflight\_lo} serves
as a \emph{short-term} inflight bound
and is therefore reset at the end of
the bandwidth-probing period. 
In summary, at any point in time,
the congestion-window size of a BBRv2 flow 
is the minimum of the general BBR congestion
window of two BDP, the short-term 
bound \texttt{inflight\_lo} (if set)
and the long-term bound \texttt{inflight\_hi}
(discounted by \texttt{headroom} in cruising mode).

\subsection{Basic fluid model for BBR}
\label{sec:model:bbr:basic}
We rely on a skeleton fluid model that
captures the common properties of 
BBRv1 and BBRv2. As mentioned in the
previous section, the two versions of
BBR are mostly similar regarding
the estimation of the minimum
RTT given by~\texttt{RTprop},
which we represent with
variable~$\tau^{\min}_i(t)$
for the~\texttt{RTprop} estimate
of agent~$i$ at time~$t$.
The variable~$\tau^{\min}_i$ is
continuously adjusted downwards
upon encountering smaller RTTs:
\begin{equation}
    \dot{\tau}^{\min}_i = -\Gamma\big(\tau^{\min}_i(t) - \tau_i(t-\propdp{})\big)
\end{equation}
where~$\Gamma(v)$
is a differentiable function
approximating the ReLU
function~$\max(0, v)$. Such a function can be constructed
using the sigmoid function from~\cref{eq:sigmoid}:
\begin{equation}
    \Gamma(v) = v \cdot \sigma(v).
\end{equation}

To describe that BBR is in \texttt{ProbeRT}
state, we use a \emph{discrete mode 
variable}~$m^{\mathrm{prt}}_i$,
which is 1 if BBR is in \texttt{ProbeRT}
state and 0 otherwise. In both BBR
versions, the \texttt{ProbeRT} mode
is switched on or off
upon time-out of the \texttt{ProbeRT} 
timer~$t^{\mathrm{prt}}_i$:
\begin{equation}
    \Delta m^{\mathrm{prt}}_i = \sigma\big(t^{\mathrm{prt}}_i-T^{\mathrm{prt}}_i\big) \cdot \Big( (1-m^{\mathrm{prt}}_i) - m^{\mathrm{prt}}_i\Big)
    \label{eq:model:bbr:mprt}
\end{equation} where~$T^{\mathrm{prt}}_i$ 
is the time
period between entries and exits of the 
\texttt{ProbeRT} state for agent~$i$.
Note that~\cref{eq:model:bbr:mprt} represents
an update rule for simulations rather 
than a differential equation, as~$m^{\mathrm{prt}}_i$
is discrete.
The two  time-related variables in~\cref{eq:model:bbr:mprt}
behave as follows:
\begin{equation}
    T^{\mathrm{prt}}_i = m^{\mathrm{prt}}_i \cdot 0.2 + (1-m^{\mathrm{prt}}_i) \cdot 10
\end{equation}
\begin{equation}
    \dot{t}^{\mathrm{prt}}_i = 1 - \sigma(t^{\mathrm{prt}}_i-T^{\mathrm{prt}}_i) \cdot
    t^{\mathrm{prt}}_i - \sigma\big(\tau^{\min}_i - \tau_i(t - \propd{})\big) \cdot t^{\mathrm{prt}}_i
\end{equation}

In \texttt{ProbeRT} mode, the sending
rate is limited by a version-dependent
inflight limit~$w^{\mathrm{prt}}_i(t)$:
\begin{equation}
    x_i = m^{\mathrm{prt}}_i \cdot \frac{w^{\mathrm{prt}}_i}{\tau_i} 
    - (1-m^{\mathrm{prt}}_i) \cdot x^{\mathrm{pbw}}_i
\end{equation}
where the \texttt{ProbeBW} sending
rate~$x^{\mathrm{pbw}}_i$ follows the relevant constraint 
(congestion window~$w_i^{\mathrm{pbw}}$ or pacing rate~$x^{\mathrm{pcg}}_i$):
\begin{equation}
    x_i^{\mathrm{pbw}} = 
    \min\big(\frac{w_i^{\mathrm{pbw}}}{\tau_i}, x^{\mathrm{pcg}}_i\big)
    \label{eq:model:bbr:x_pbw}
\end{equation}

Similar to the \texttt{ProbeRT} state,
we also introduce the two time-related
variables~$T^{\mathrm{pbw}}_i$ 
and~$t^{\mathrm{pbw}}_i$ for 
the~\texttt{ProbeBW} state, where
$T^{\mathrm{pbw}}_i$ is the duration
of a \texttt{ProbeBW} period 
and~$t^{\mathrm{pbw}}_i$ is the 
time within the current period.
While~$T^{\mathrm{pbw}}_i$ is fully
version-dependent,~$t^{\mathrm{pbw}}_i$
grows with time and
is reset to 0 when 
exceeding the period duration
for both BBR versions:
\begin{equation}
    \dot{t}^{\mathrm{pbw}}_i =
    1 - \sigma(t^{\mathrm{pbw}}_i - T^{\mathrm{pbw}}_i) \cdot t^{\mathrm{pbw}}_i 
\end{equation}

In the \texttt{ProbeBW} state,
the bottleneck-bandwidth estimation
is based on measurements of the
delivery rate~$x_i^{\mathrm{dlv}}$
(with link~$\ell$ being the bottleneck link
of agent~$i$):
\begin{equation}
    x_i^{\mathrm{dlv}} = \frac{\rate{t-\propdp{}}}{\arriv{t-\propdb{}}} \cdot \begin{cases}
        \lcap{} & \text{if } q_{\ell}(t-\propdb{}) > 0\\
        \arriv{t-\propdb{}} & \text{otherwise}
    \end{cases}
    \label{eq:bbr:x_dlv}
\end{equation}
We accommodate the recorded maximum
delivery rate~$x^{\max}_i(t)$
per \texttt{ProbeBW} period as follows:
\begin{equation}
    \dot{x}^{\max}_i =  \Gamma(x_i - x^{\max}_i) - \sigma(0.01 - t^{\mathrm{pbw}}_i) \cdot x^{\max}_i
\end{equation} where the second term
provokes
a reset of~$x^{\max}_i$ in the first
ten milliseconds of the period.
The mechanism for adjusting the 
bottleneck-bandwidth estimate~$x^{\mathrm{btl}}_i$
(corresponding to \texttt{BtlBw})
to~$x^{\max}_i$ is specific to each
BBR version.

Finally, we choose the following
natural formulation to model the inflight
volume~$v_i(t)$:
\begin{equation}
    \dot{v}_i = x_i - x^{\mathrm{dlv}}_i
\end{equation}

\subsection{BBRv1 Fluid Model}
\label{sec:model:bbr:bbrv1}

Given the basic BBR fluid-model
framework, the biggest
challenge in modelling BBRv1 is
to model the randomized
probing behavior with varying pacing rates.
As described in~\cref{sec:model:bbr:description},
BBRv1 proceeds in bandwidth-probing
periods that are 8 phases long,
where each phase has a duration 
of~$\tau^{\min}_i$, i.e., 
$T^{\mathrm{pbw}}_i = 8\cdot\tau^{\min}_i$.
The bottleneck-bandwidth 
estimate~$x^{\mathrm{btl}}_i$ is
updated to the maximum
delivery rate~$x^{\max}_i$ at the
end of the period, which we formalize
by the following
equation:
\begin{equation}
    \dot{x}^{\mathrm{btl}}_i = 
    \sigma\big(t^{\mathrm{pbw}}_i - T^{\mathrm{pbw}}_i + 0.01\big) \cdot \big(x^{\max}_i - x^{\mathrm{btl}}_i\big)
    \label{eq:model:bbr:1:xbtl}
\end{equation}

\begin{figure}
    \begin{subfigure}[b]{0.27\linewidth}
        \centering
        \includegraphics[width=\linewidth]{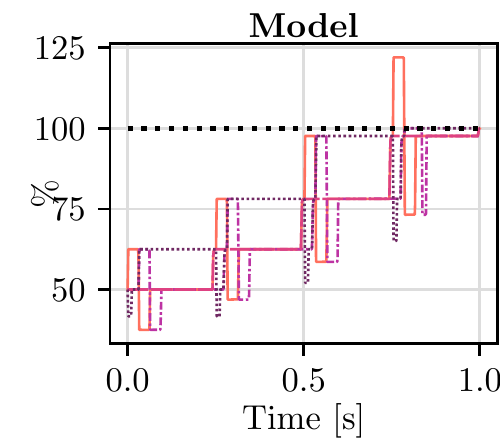}
        \caption{BBRv1.}
        \label{fig:model:bbr:1}
    \end{subfigure}
    \begin{subfigure}[b]{0.72\linewidth}
        \centering
        \includegraphics[width=\linewidth]{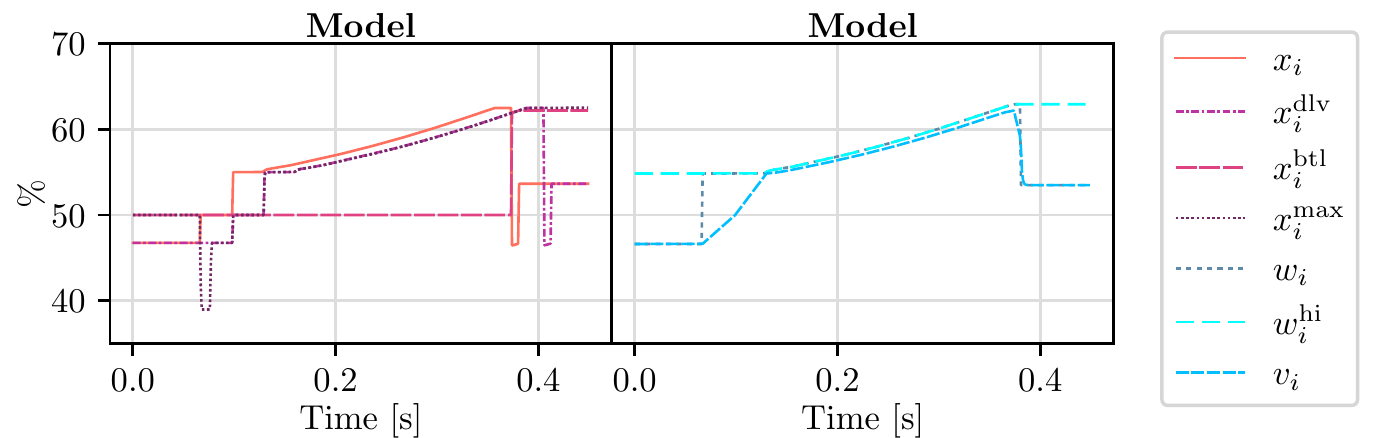}
        \caption{BBRv2 (left: rate, right: inflight limits).}
        \label{fig:model:bbr:2}
    \end{subfigure}
    \vspace{-10pt}
    \caption{Visualization of BBR fluid-model variables link capacity normalized to 100\%, single flow)}
    \label{fig:model:bbr}
\end{figure}

In general, BBRv1 prescribes a
pacing rate~$x^{\mathrm{pcg}}_i$
equal to~$x^{\mathrm{btl}}_i$
in each phase, but increases
$x^{\mathrm{pcg}}_i$ to $\nicefrac54\cdot x^{\mathrm{btl}}_i$
in one randomly chosen phase
and decreases it to 
$\nicefrac34\cdot x^{\mathrm{btl}}_i$
in the subsequent phase.
To restrict a given behavior to
a certain phase~$\phi \in \{0, ..., 7\}$, 
we introduce the following \emph{pulse} 
function~$\Phi$, which is 1
if BBRv1 is in phase~$\phi$
and 0 otherwise:
\begin{equation}
    \Phi_i(t, \phi) = \sigma\big(t^{\mathrm{pbw}}(t) - \phi\cdot\tau^{\min}_i\big) \cdot \sigma\big((\phi+1)\cdot\tau^{\min}_i - t^{\mathrm{pbw}}\big)
\end{equation}
This pulse function allows to model
the pacing behavior of an agent~$i$
that employs the augmented
pacing rate in phase~$\phi_i$:
\begin{equation}
    x^{\mathrm{pcg}}_i = x^{\mathrm{btl}}_i \cdot
    \big(1 + \tfrac14 \cdot \Phi_i(t, \phi_i) 
    -\tfrac14 \cdot \Phi_i(t, \phi_i+1)\big)
\end{equation}
In the implementation of
BBRv1, the phase~$\phi_i$ is randomly
chosen from $\{0, ..., 6\}$
every time BBRv1 switches
from~\texttt{ProbeRT} state back
to \texttt{ProbeBW} state.
Since such randomness is incompatible
with the determinism of fluid models, 
we mimic the randomness of~$\phi_i$
by choosing it as~$i \bmod 6$, where
we assume the agent identifier~$i$
to be a natural number. This agent-dependent
choice of~$\phi_i$ desynchronizes
the pacing-rate variation of agents~$i$
on paths with equal RTT, which is the
central goal of the randomization,
without sacrificing the determinism
of the fluid model. The interplay
of BBRv1 variables in pacing-based
mode is visualized 
in~\cref{fig:model:bbr:1}.

The basic BBR fluid-model
allows a straightforward integration
of the state-dependent inflight limits
of BBRv1:
\begin{equation}
    w^{\mathrm{prt}}_i = 4
    \hspace{50pt}
    w^{\mathrm{pbw}}_i = 2 \cdot \overline{w}_i = 2 \cdot x^{\mathrm{btl}}_i \cdot \tau^{\min}_i
\end{equation}
where~$\overline{w}_i$ denotes
the BDP estimated by agent~$i$.

\subsection{BBRv2 Fluid Model}
\label{sec:model:bbr:bbrv2}

BBRv2 mostly differs from BBRv1
with regard to the structure of
the bandwidth-probing phase
in several ways.

First, a bandwidth-probing period
is considerably longer
than in BBRv1: The duration of 
bandwidth-probing periods in BBRv2
is given by the minimum of 62 estimated
RTTs and a random value between 2 and 3 seconds.
This randomness in BBRv2 poses a
similar challenge in modelling as 
the randomness in BBRv1,
such that we again use an approach
based on the agent identifier
to achieve the central goal of agent
desynchronization without sacrificing determinism:
\begin{equation}
    T^{\mathrm{pbw}}_i = 
    \min\big(63\cdot \tau^{\min}_i, 2 + \frac{i}{N}\big)
\end{equation}

Second, the behavior in the bandwidth-probing phases of
BBRv2 differs from BBRv1.
To model the BBRv2 phases, we
introduce two additional mode variables,
namely~$m^{\mathrm{dwn}}_i(t)$, which indicates
whether agent~$i$ is attempting to reduce
its inflight at time~$t$, 
and~$m^{\mathrm{crs}}_i(t)$, which indicates
whether agent~$i$ is cruising at
time~$t$.
The mode variable~$m^{\mathrm{dwn}}_i$ affects the 
pacing rate~$x^{\mathrm{pcg}}_i$ as follows:
\begin{equation}
    x^{\mathrm{pcg}}_i = x^{\mathrm{btl}}_i \cdot \Big(1 + \tfrac14 \cdot \sigma(t^{\mathrm{pbw}}_i - \tau^{\min}_i) \cdot
    (1 - m^{\mathrm{dwn}}_i) - \tfrac14 \cdot m^{\mathrm{dwn}}_i\Big)
\end{equation} where~$m^{\mathrm{dwn}}_i$
increases the pacing rate to $\nicefrac54 \cdot x^{\mathrm{btl}}_i$ if $m^{\mathrm{dwn}}_i = 0$ 
(and one RTT has
passed in the bandwidth-probing period), and
decreases the pacing rate to $\nicefrac34 \cdot x^{\mathrm{btl}}_i$ if~$m^{\mathrm{dwn}}_i = 1$.

Third, while we modelled phase transitions in BBRv1
as purely dependent on time~$t^{\mathrm{pbw}}_i$,
the phase transitions in BBRv1 are triggered
by probing observations. In particular,
the inflight-reducing mode~$m^{\mathrm{dwn}}_i$
is activated if the inflight~$v_i$
exceeds~$\nicefrac54\cdot\overline{w}_i$
or loss~$p_{\pi_i}$ exceeds 2\%,
and is disabled once the reduced
pacing rate has managed to reduce
the inflight~$v_i$ to
the draining target~$w_i^- = \min(\overline{w}_i, 0.85 \cdot w_i^{\mathrm{hi}})$:
\begin{equation}
    \begin{split}
        \Delta m^{\mathrm{dwn}}_i 
    =\ &(1-m^{\mathrm{crs}}_i) \cdot
    (1-m^{\mathrm{dwn}}_i) \cdot
    \sigma(t^{\mathrm{pbw}}_i - \tau^{\min}_i) \\ 
    &\cdot
   \min\big(\sigma(v_i - \nicefrac54\cdot\overline{w}_i) +
    \sigma(p_{\pi i} - 0.02),\ 1\big)\\
    &- m^{\mathrm{dwn}}_i \cdot \sigma\Big(w_i^- - v_i\Big)
    \label{eq:model:bbr:mdwn}
    \end{split}
\end{equation} where~$w^{\mathrm{hi}}$ 
is the variable accommodating the
long-term bound~\texttt{inflight\_hi}
in our model.
Moreover, the disabling of~$m^{\mathrm{dwn}}_i$
automatically leads to the activation
of~$m^{\mathrm{crs}}_i$, which is then
disabled again when a new bandwidth-probing
period starts:
\begin{equation}
    \Delta m^{\mathrm{crs}}_i =
    -\Delta m^{\mathrm{dwn}}_i - \sigma(t^{\mathrm{pbw}}_i - T^{\mathrm{pbw}}_i) \cdot m^{\mathrm{crs}}_i
\end{equation}

The fourth difference concerns the adjustment of
the bottle-neck-bandwidth 
estimate~$x^{\mathrm{btl}}_i$.
In BBRv2, $x^{\mathrm{btl}}_i$ is
adjusted to the maximum
delivery rate from the last two
probing periods when the
inflight-growing phase has stopped:
\begin{equation}
    \dot{x}^{\mathrm{btl}}_i =
    m^{\mathrm{dwn}}_i \cdot \Big(\max\big(x^{\max}_i,\ x^{\max}_i(t-T^{\mathrm{pbw}})\big) - x^{\mathrm{btl}}_i\Big)
\end{equation}

Fifth, BBRv2 operates with another two
additional state variables, namely
\texttt{inflight\_hi} and \texttt{inflight\_lo},
which we accommodate in our fluid model
with~$w^{\mathrm{hi}}_i$ and~$w^{\mathrm{lo}}_i$,
respectively. The upper inflight
bound~$w^{\mathrm{hi}}$ is exponentially
adjusted upwards when it represents 
the relevant constraint on the sending rate
($v_i = w^{\mathrm{hi}}_i$)
during the aggressive probing phase
and no excessive loss occurs.
In contrast, \texttt{inflight\_hi} 
is reduced by a multiplicative
decrease of 30\% if encountering
loss exceeding~2\%. To be precise,
the BBRv2 implementation applies 
this multiplicative decrease at 
most once per bandwidth-probing period.
We approximate this behavior
with a reduced multiplicative decrease
in presence of excessive loss:
\begin{equation}
    \begin{split}
        \dot{w}^{\mathrm{hi}}_i = &(1 - m^{\mathrm{crs}}_i) \cdot \sigma(t^{\mathrm{pbw}}_i - \tau^{\min}_i) \cdot \sigma(v_i - w^{\mathrm{hi}})\\ 
        &\cdot 2^{t^{\mathrm{pbw}}_i/\tau^{\min}_i} - \sigma(p_{\pi_i} - 0.02) \cdot \frac{0.3}{\tau^{\min}_i} \cdot w^{\mathrm{hi}}_i
    \end{split}
    \label{eq:bbr2:w_hi}
\end{equation}
Outside of cruising mode, the lower inflight~
bound~$w^{\mathrm{lo}}_i$
is unset (which we represent with an
assimilation to $w_i^-$).
In cruising mode, $w^{\mathrm{lo}}_i$ is
also decreased by 30\%
per RTT upon encountering loss:
\begin{equation}
        \dot{w}^{\mathrm{lo}}_i = (1- m^{\mathrm{crs}}_i) \cdot
    (w_i^- - w^{\mathrm{lo}}_i) - m^{\mathrm{crs}}_i \cdot \sigma(p_{\pi_i}) \cdot \frac{0.3w^{\mathrm{lo}}_i}{\tau^{\min}_i} \cdot 
\end{equation}
In summary, the congestion-window size in \texttt{ProbeBW}
state is given as follows in BBRv2:
\begin{equation}
    w^{\mathrm{pbw}}_i = \min\Big(2\cdot\overline{w}_i,\  (1 - m^{\mathrm{crs}}_i) \cdot w^{\mathrm{hi}}_i + m^{\mathrm{crs}}_i \cdot w_i^{\mathrm{lo}}\Big)
    \label{eq:bbr2:w_pbw}
\end{equation}

A final difference between BBRv1 and BBRv2
concerns the congestion-window size
in \texttt{ProbeRT} mode. Instead of
using a fixed congestion-window size 
of 4 segments, BBRv2 cuts the congestion window
to half the estimated BDP in this mode:
\begin{equation}
    w^{\mathrm{prt}}_i = \frac{\overline{w}_i}{2}
\end{equation}

The interplay of the variables in the BBRv2
fluid model is visualized by means of an example
in~\cref{fig:model:bbr:2}.
%%%%%%%%%%%%%%%%%%%%%%
\section{Experimental Validation}
\label{sec:experiments}

In this section, we experimentally
validate our BBR fluid model,
% presented
% in the preceding section, 
building
on the network emulator mininet~\cite{lantz2010network}. 
% We perform this validation as described
% in~\cref{sec:experiments:setup} and
% in terms of both 
% traces (cf.~\cref{sec:experiments:traces}) 
% and aggregate 
% metrics (cf.~\cref{sec:experiments:aggregate}).

\subsection{Validation Set-up}
\label{sec:experiments:setup}

\subsubsection{Model-Based Computations}

The fluid models have been
implemented using NumPy~\cite{numpy}. 
We will make this 
implementation available. Differential
equations are solved with the method
of steps~\cite[\S 1.1.2]{erneux2009applied}
with a step size of \SI{10}{\us}.

\subsubsection{Experiments}

To compare the model output with
implementation behavior,
we perform experiments using the
network emulator 
mininet~\cite{lantz2010network}.
% The mininet implementation uses
% additional network interfaces
% configured with NetEm~\cite{netem} 
% to emulate links
% with the desired rates, delays 
% and queuing disciplines. 
In mininet, we
use OvS to emulate 
switches~\cite{openvswitch}.
The emulated hosts send traffic
by using iPerf~\cite{iperf}.
The complete infrastructure
for measurements and emulation
will be published.
All experiments are 
run with an Intel Core 
Intel Xeon E5-2695 v4 CPU.

\subsubsection{Topology}

\begin{figure}
    \centering
    \begin{tikzpicture}[
	switchnode/.style={circle, draw=black!60, shading=radial,outer color={rgb,255:red,137;green,207;blue,240},inner color=white, thick, minimum size=8mm},
	hostnode/.style={circle, draw=black!60, shading=radial,outer color={rgb,255:red,255;green,153;blue,102},inner color=white, thick, minimum size=8mm},
	]
	
	\node[hostnode]	 (a1)	  at (-2.75, 1.2)	{$\boldsymbol{a_1}$}; 
	\node[hostnode]	 (a2)	  at (-2.75, 0.5)	{$\boldsymbol{a_2}$};
	
	\node (dots) at (-2.75, 0) {...};
	
	\node[hostnode]    (a10)    at (-2.75, -0.5) {$\boldsymbol{a_{N}}$};
	
	\node[switchnode]    (R1)     at (0, 0.35) {$\boldsymbol{S}$};
	
	\node[hostnode]    (S1)     at (2.75, 0.35) {$\boldsymbol{a_{\mathrm{d}}}$};
	
	\draw (a1.east)  to (R1.west);
	\draw (a2.east)  to (R1.west);
	\draw (a10.east) to (R1.west);
	
	\draw (R1.east) to (S1.west);
	
	\node (l1)  at (-1.5, 1.0) {$\ell_1$};
	\node (l2)  at (-1.85, 0.2) {$\ell_2$};
	\node (l10) at (-1.5, -0.5) {$\ell_{N}$};
	
	\node (l) at (1.375, 0.5) {$\ell$};

\end{tikzpicture}
    \vspace{-10pt}
    \caption{Dumbbell topology.}
    \label{fig:topology}
    \vspace{-10pt}
\end{figure}
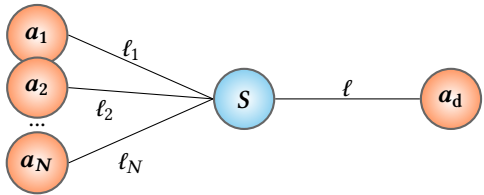

As usual in the 
literature~\cite{ware2019modeling,song2021understanding,gomez2020performance,hock2017experimental,kfoury2020emulation,nandagiri2020bbrvl,scholz2018towards},
we consider the dumbbell topology
in~\cref{fig:topology}.
In this topology, $N$ agents~$a_i$, 
$i\in\{1,...,N\}$, communicate with 
a destination host~$a_{\mathrm{d}}$
via a switch~$S$. In all
paths, the shared link~$\ell$ between
switch~$S$ and destination 
host~$a_{\mathrm{d}}$ constitutes
the bottleneck link. 
The links~$\ell_i$, 
$i \in \{1,...,N\}$, which connect
the individual senders to switch~$S$,
are never saturated and therefore
do not affect the sending rates.
The propagation delays
of these non-shared links
are heterogeneous (randomly selected
from a given range) such that 
the individual senders experience
different RTTs.
Switch~$S$ is equipped
with a buffer, the size of which
is measured in
bandwidth-delay product~(BDP) of 
the bottleneck link~$\ell$.

\subsection{Validation of Trace Results}
\label{sec:experiments:traces}

\begin{figure*}
    %\vspace{-10pt}
    \begin{subfigure}[b]{0.4\linewidth}
        \centering
        \includegraphics[width=\linewidth]{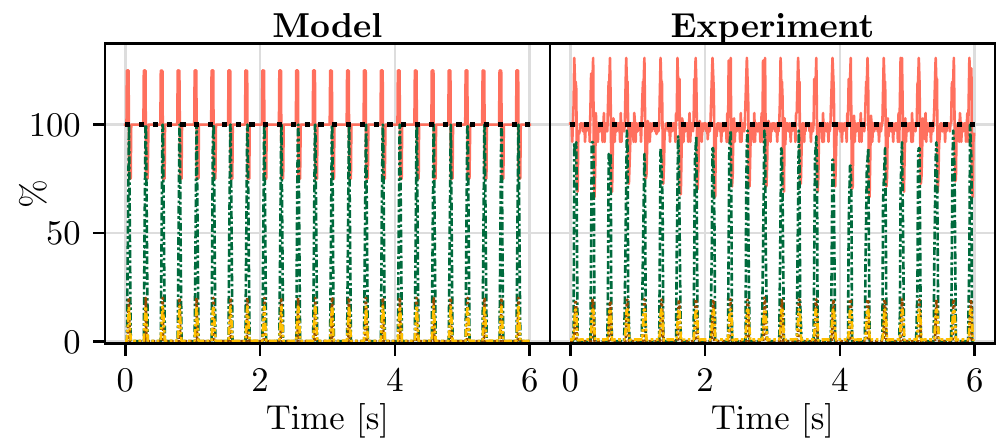}
        \vspace{-15pt}
        \caption{Drop-tail}
        \label{fig:validation:traces:bbr:droptail}
    \end{subfigure}
    \begin{subfigure}[b]{0.55\linewidth}
        \centering
        \includegraphics[width=\linewidth]{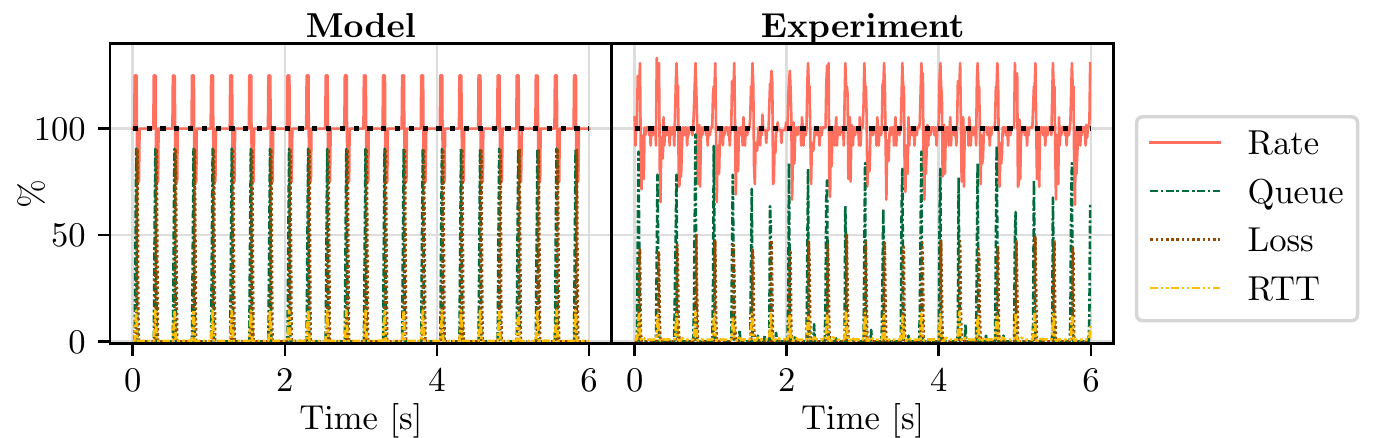}
        \vspace{-15pt}
        \caption{RED}
        \label{fig:validation:traces:bbr:red}
    \end{subfigure}
    \vspace{-15pt}
    \caption{BBRv1 trace validation}
    \label{fig:validation:traces:bbr}
    \vspace{-5pt}
\end{figure*}

\begin{figure*}
    %\vspace{-10pt}
    \begin{subfigure}[b]{0.4\linewidth}
        \centering
        \includegraphics[width=\linewidth]{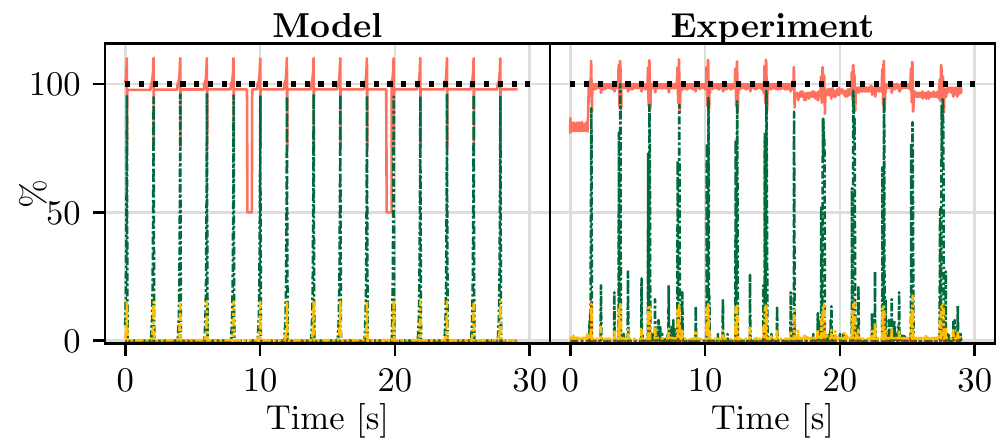}
        \vspace{-15pt}
        \caption{Drop-tail}
        \label{fig:validation:traces:bbr2:droptail}
    \end{subfigure}
    \begin{subfigure}[b]{0.55\linewidth}
        \centering
        \includegraphics[width=\linewidth]{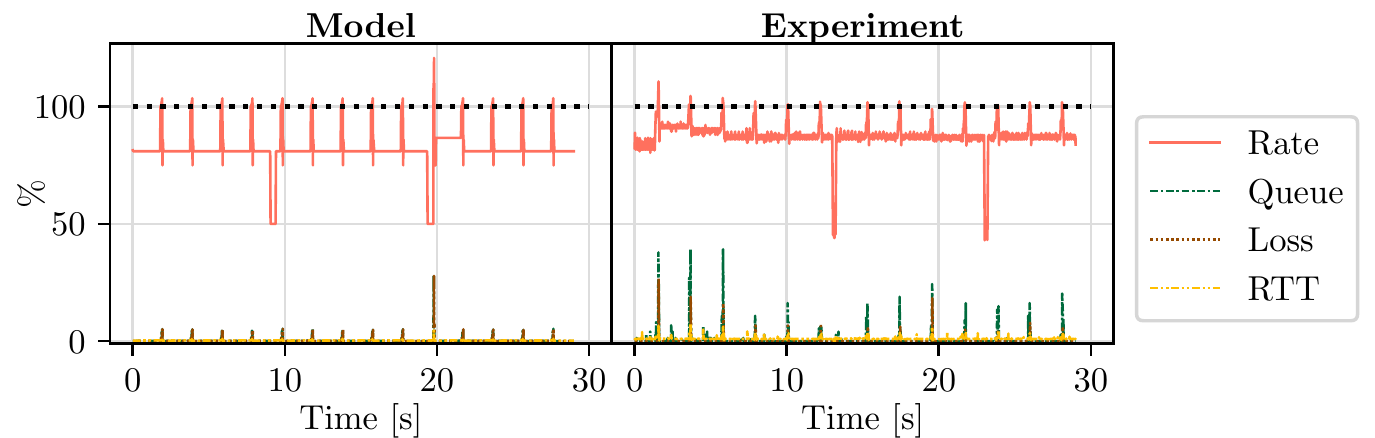}
        \vspace{-15pt}
        \caption{RED}
        \label{fig:validation:traces:bbr2:red}
    \end{subfigure}
    \vspace{-15pt}
    \caption{BBRv2 trace validation}
    \label{fig:validation:traces:bbr2}
    \vspace{-5pt}
\end{figure*}

Using the validation set-up
described 
in~\cref{sec:experiments:setup},
we first verified
traces as predicted
by the fluid models regarding their
similarity with traces
from experimental measurements.
The concrete network setting
in this trace validation 
included a single sender,
a bottleneck link~$\ell$ with a rate
of $\lcap{} = \SI{100}{Mbps}$ 
(as recommended by 
mininet~\cite{mininet_intro})
and with a propagation 
delay of~$\propd{} = \SI{10}{\ms}$,
a non-bottleneck link~$\ell_1$
with a delay 
of~$d_{\ell_1} = \SI{5.6}{\ms}$,
and a switch buffer of 1 BDP.

\cref{fig:validation:traces:bbr,fig:validation:traces:bbr2}
visualize the comparison of
the thus obtained traces
for the BBRv1 and BBRv2,
respectively, where each CCA
was tested under both a drop-tail
and a RED queuing discipline
(A validation of Reno and CUBIC
can be found in~\cref{sec:model-cc}). 
All measurements have been normalized:
The sending rate is given in percent
of the bottleneck-link rate,
the queue in percent
of buffer volume, loss 
in percent of traffic volume,
the RTT as relative excess
delay, and the
congestion window in
percent of the path 
BDP. 
The comparisons highlight both
important commonalities and 
differences between the model
and the experiments.

\textbf{Similarities.} The
fluid models are highly predictive
regarding the rate patterns
over time. 
In addition, the fluid models correctly 
capture that the loss-sensitive BBRv2
lead to considerably smaller loss
(barely visible) than BBRv1, which is
insensitive to loss.
Finally, the fluid model correctly predicts
that the sending rate of the loss-sensitive BBRv2 
barely exceeds the bottleneck rate under RED, 
while RED has no impact on the loss-insensitive BBRv1.
As a result, the smaller
buffer usage of BBRv1 under RED is also reflected
in the model, although the difference
between RED and drop-tail is more pronounced
in the model.

\textbf{Difference: RED idealization.} 
The difference above is due to the 
idealization of RED:
In the model, the queue size affects
the loss probability instantly;
in reality, RED relies
on outdated and averaged measurements
of the queue size, causing some lag between
queue build-ups and loss surges.

\textbf{Difference: \texttt{ProbeRTT} state in BBRv2.}
In the model, the BBRv2 flow
regularly manages to drain the queue,
therefore discovers the propagation delay
early, and cannot detect a lower RTT 
afterwards; hence, it enters the \texttt{ProbeRTT} state
every 10 seconds. Thanks to the low
queue build-up, the BBRv2 flow in
the RED experiment also discovers
the propagation delay.
However, the BBRv2 flow in the
drop-tail experiment 
can regularly decrease its
\texttt{RTprop} estimate because of
the small, but persistent and fluctuating queue,
and therefore never enters 
the \texttt{ProbeRTT} state.
% The fluctuating queue is never fully
% drained because
% the BBRv2 flow fully saturates
% the link capacity and
% includes random fluctuations,
% which are by design not represented
% in the fluid model.

In summary, the
fluid models capture the differences among
CCAs and queuing disciplines
with high accuracy, 
especially qualitatively
and to a lesser degree also quantitatively.

\subsection{Validation of Aggregate Results}
\label{sec:experiments:aggregate}

The trace validations in the previous
subsection indicate that the presented CCA fluid models
yield reasonable predictions for single senders. 
The more important question, however, is whether these fluid models
can acceptably predict 
network-performance metrics given interacting senders.
To test the fluid models
in this metric-oriented aspect, 
we compare aggregate results from 
model computations and experiments
for a wide variety of network parameters,
in particular with respect to
Jain fairness~(\cref{fig:validation:10ms:fairness}),
% \footnote{Given rates~$x_i$, $i\in[N]:=\{1,...,N\}$,
% of different senders, the Jain fairness index
% is computed 
% as $(\sum_{i\in[N]}x_i)^2/(N\cdot\sum_{i\in[N]}x_i^2)$.},
packet loss~(\cref{fig:validation:10ms:loss}),
buffer occupancy~(\cref{fig:validation:10ms:queuing}),
bottleneck-link utilization~(\cref{fig:validation:10ms:utilization})
and jitter, i.e., packet-delay 
variation~(\cref{fig:validation:10ms:jitter}).
All metrics were obtained from the aggregation
of 5-second traces, where the experiment results
are averaged over 3 runs. In contrast, fluid models are deterministic and
do not require averaging.
The network setting was based on the
topology in~\cref{fig:topology},
$N= 10$ senders, a bottleneck-link rate of
$\lcap{} = \SI{100}{Mbps}$, a bottleneck-link propagation 
delay~$\propd{} = \SI{10}{\ms}$ and total
RTTs randomly selected between 30 and 40ms.
For heterogeneous CCAs,
each CCA was employed by $N/2 = 5$ senders.
To strengthen our validation, we conduct
the same analysis for shorter delays,
which confirms our results (cf.~\cref{sec:aggregate-validation:5ms}).

As in the trace validation, the comparisons
reveal that the model predictions and the 
experiment results have striking similarities
as well as notable differences.
In the following, we will discuss each
metric separately.

\subsubsection{Fairness}
\label{sec:experiments:aggregate:fairness}

\begin{figure*}
    \begin{subfigure}[b]{0.4\linewidth}
        \centering
        \includegraphics[width=\linewidth]{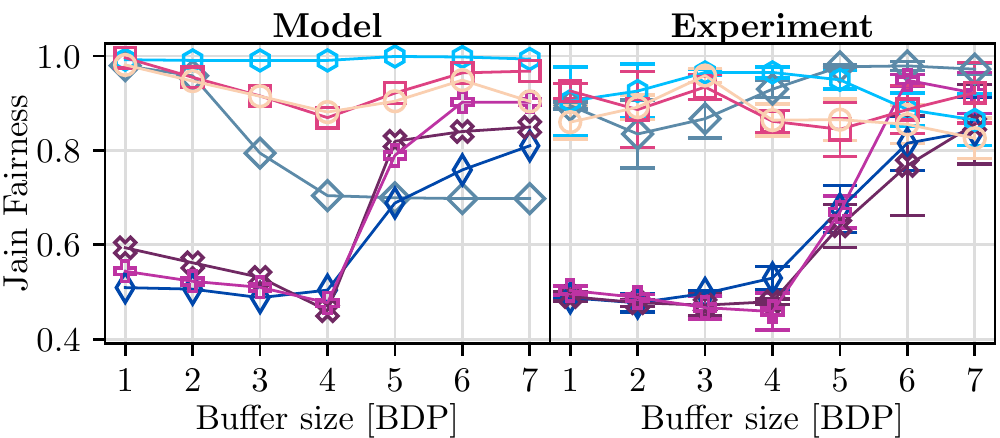}
        \caption{Drop-tail}
        \label{fig:validation:10ms:fairness:droptail}
    \end{subfigure}
    \begin{subfigure}[b]{0.55\linewidth}
        \centering
        \includegraphics[width=\linewidth]{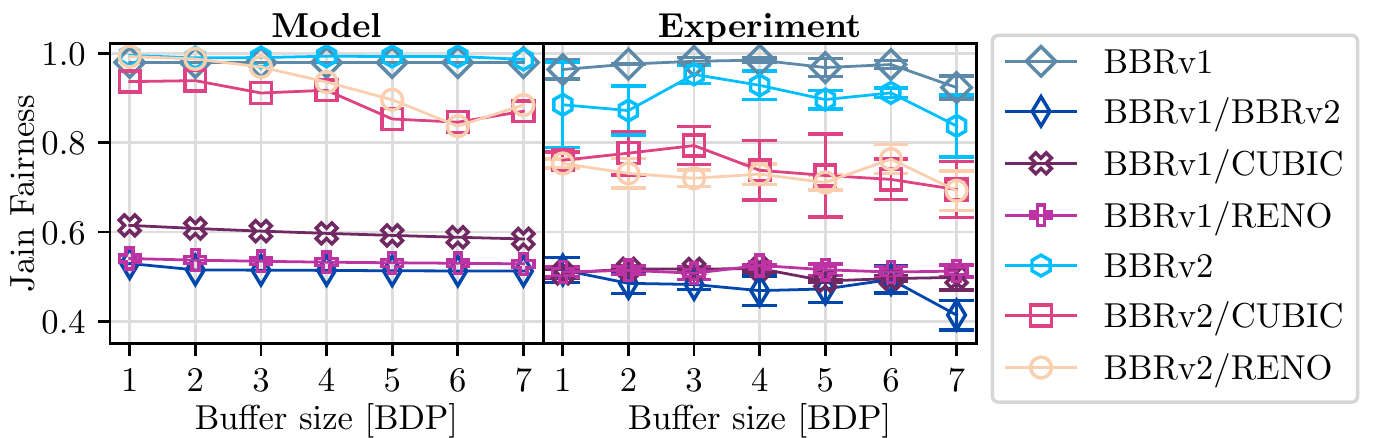}
        \caption{RED}
        \label{fig:validation:10ms:fairness:red}
    \end{subfigure}
    \vspace{-15pt}
    \caption{Fairness validation}
     \label{fig:validation:10ms:fairness}
\end{figure*}

Regarding fairness~(cf.~\cref{fig:validation:10ms:fairness}),
we first observe that the least fairness arises when
a loss-sensitive CCA (Reno, CUBIC or BBRv2) is in competition
with BBRv1 in shallow buffers, which has already been well documented
in previous research~\cite{scholz2018towards,ware2019modeling}.
This unfairness is the result of the loss insensitivity
of BBRv1, which maintains its rate despite loss
while loss-sensitive CCAs practically stop sending
in reaction to the loss caused by BBRv1.
Starting at buffer sizes from 4 BDP, however,
the fairness in these settings increases
for two reasons. First, these large buffers reduce
the occurrence of loss, which prevents the back-off
of loss-sensitive CCAs. Second, in large buffers, 
the inflight limit given by the congestion window of BBRv1
restricts the sending rate of BBRv1 
and allows competing flows to obtain a higher share of bandwidth
than in shallow buffers. Given a RED queue, however,
the fairness of BBRv1 towards loss-sensitive CCAs is
consistently low because RED (1)
increases loss and (2) restricts the buffer build-up
such that the inflight of the BBRv1 flows is substantially
below their inflight limit.

The fairness issues of BBRv1 appear to
have been largely resolved with BBRv2, as
both the fluid model and the experiment results show.
However, BBRv2 is still unfair towards loss-based CCAs
in RED buffers, where the higher loss sensitivity
of loss-based CCAs is revealed.

One substantial difference between the fluid-model
predictions and the experiment results is the 
decreasing fairness of BBRv1 in deep drop-tail
buffers, which only appears in the fluid model.
The fluid model reveals 
the RTT unfairness of BBRv1, which has indeed
been experimentally confirmed~\cite{scholz2018towards,ware2019modeling}, 
although for higher RTT differences than used in our
network setting. This RTT unfairness stems from
the inflight limit of BBRv1, which becomes relevant in deep buffers: 
Since flows with a lower RTT estimate a lower BDP and hence
maintain a smaller congestion window, lower-RTT flows
are more severely restricted by the inflight limit.
Moreover, these differences are increased over time
because flows with a higher rate also
measure a higher maximum delivery rate, which
affects the bottleneck-bandwidth estimate
in the estimated BDP. This effect can be
analytically revealed by means of~\cref{eq:bbr:x_dlv}
for the delivery rate~$x^{\mathrm{dlv}}$:
Given an existing queue~$q_{\ell} > 0$,
any reduction in total arrival rate~$y_{\ell}$ (e.g., because of
one sender~$j$ decreasing its pacing rate~$x^{\mathrm{pcg}}_j$
to $\nicefrac34\cdot x^{\mathrm{pcg}}_j$) results in
a larger absolute~$x^{\mathrm{dlv}}$
for senders with a high rate~$x$.
This representation of the delivery
rate idealizes the
noisy relationship between sending rates
and delivery rates in real-world buffers.
Hence, the described effect only appears
for larger RTT differences in experiments,
where the noise in delivery rates does not
eliminate the difference in sending rates
between flows with different RTTs.

In conclusion, the fluid models correctly predict
fairness effects from a qualitative perspective,
i.e., they rank CCA settings correctly according
to their fairness, and approximately also
from a quantitative perspective. Interestingly, the fluid model
also predicts RTT unfairness among BBRv1 flows in deep drop-tail
buffers, which does not appear in the corresponding experiments.
However, this RTT unfairness is a real issue in more extreme settings than tested
in this validation, i.e., for higher RTT differences 
between the senders. 
Since the role of
a fluid model is to reveal problematic CCA features
for further investigation, we argue that the exaggeration of an 
existing problem barely weakens the methodological
value of the BBR fluid model.

\subsubsection{Loss}
\label{sec:experiments:aggregate:loss}

\begin{figure*}
    %\vspace{-10pt}
    \begin{subfigure}[b]{0.4\linewidth}
        \centering
        \includegraphics[width=\linewidth]{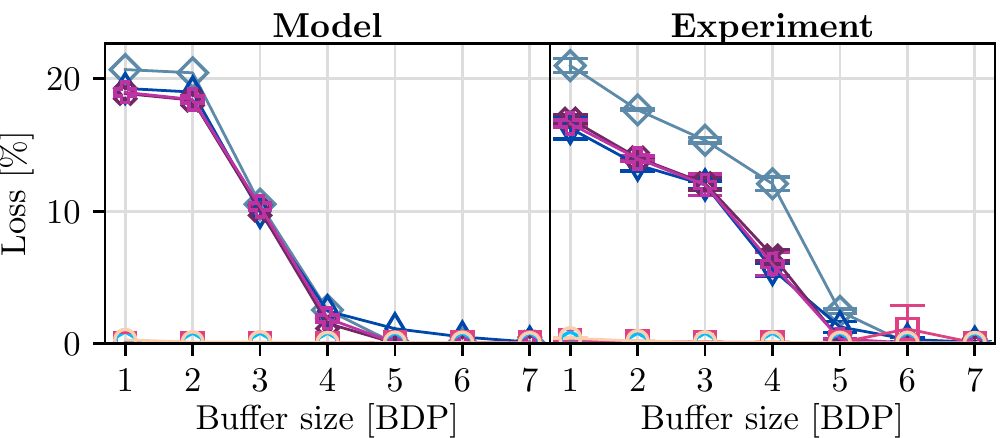}
        \caption{Drop-tail}
        \label{fig:validation:10ms:loss:droptail}
    \end{subfigure}
    \begin{subfigure}[b]{0.55\linewidth}
        \centering
        \includegraphics[width=\linewidth]{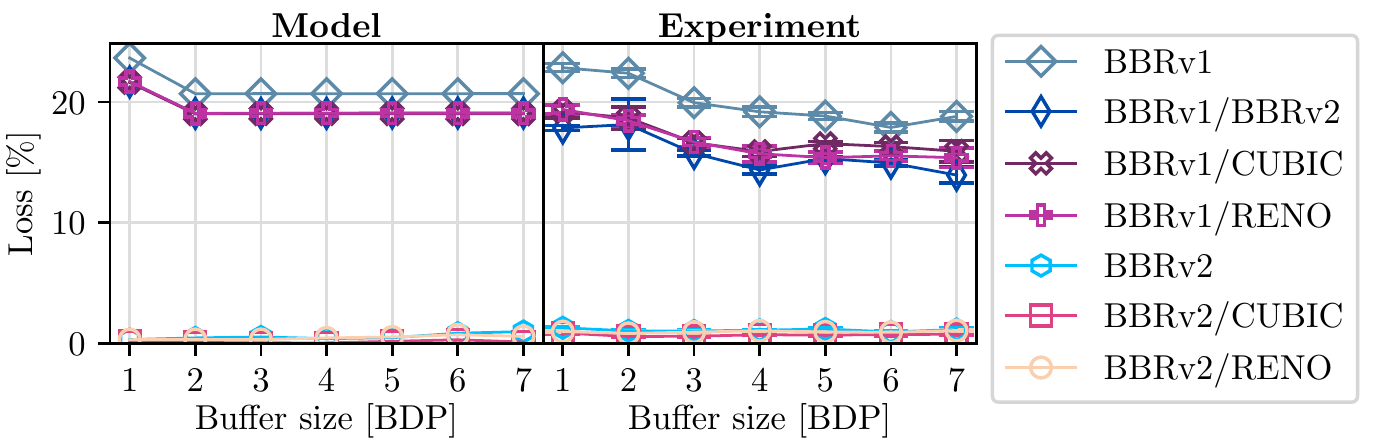}
        \caption{RED}
        \label{fig:validation:10ms:loss:red}
    \end{subfigure}\\
    \begin{subfigure}[b]{0.4\linewidth}
        \centering
        \includegraphics[width=\linewidth]{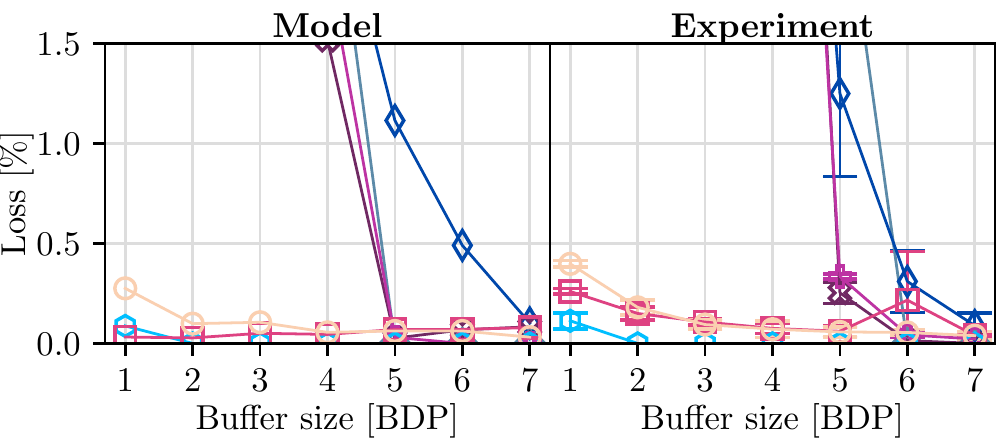}
        \caption{Drop-tail (zoomed in)}
        \label{fig:validation:10ms:loss:droptail:zoom}
    \end{subfigure}
    \begin{subfigure}[b]{0.55\linewidth}
        \centering
        \includegraphics[width=\linewidth]{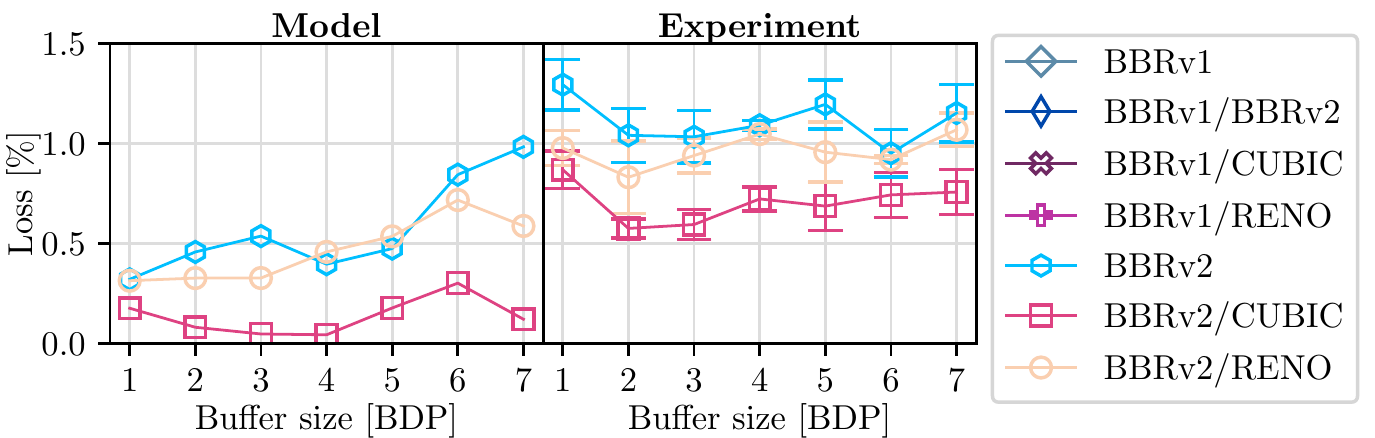}
        \caption{RED (zoomed in)}
        \label{fig:validation:10ms:loss:red:zoom}
    \end{subfigure}
    \vspace{-15pt}
    \caption{Loss validation}
    \label{fig:validation:10ms:loss}
\end{figure*}

\cref{fig:validation:10ms:loss} suggests that
fluid models are highly suitable to predict loss rates
for different CCAs, both in homogeneous settings
and heterogeneous settings and both qualitatively
and quantitatively. Our model correctly predicts
(1) that the loss rate of loss-sensitive CCAs 
(Reno, CUBIC, BBRv2 and combinations
thereof) in drop-tail buffers is below 1\% 
and goes to 0\% for increasing buffer size,
(2) that BBRv1 leads to considerable loss
of at most 20\%, where the loss rate is indirectly
proportional to the buffer size for drop-tail
queuing,
and (3) that a RED queuing discipline keeps
loss rates roughly consistent across buffer sizes.

One obvious prediction error of
the fluid model is the underestimation of
loss rates for loss-sensitive CCAs given
RED in~\cref{fig:validation:10ms:loss:red:zoom}. 
This underestimation stems
again from an idealization of the RED queue in
the model, which determines the
loss rate based on the current
queue length. In contrast, real
RED tracks the queue length
with a moving average and hence reacts
to queue build-up with delay. Since
the queue has more time to
accumulate until stabilization
by RED, the queue length
is higher than given an instantaneously
reacting RED algorithm (cf. also~\cref{fig:validation:10ms:queuing:red}).
Moreover, since the RED
dropping probability is proportional
to the queue size, the loss rate
given delayed RED 
is slightly higher than for
instantaneous RED.
However, since this underestimation
only amounts to 0.5 percentage points,
we still consider our fluid model
highly predictive with respect to loss.

\subsubsection{Queuing}
\label{sec:experiments:aggregate:queueing}

\begin{figure*}
    %\vspace{-10pt}
    \begin{subfigure}[b]{0.4\linewidth}
        \centering
        \includegraphics[width=\linewidth]{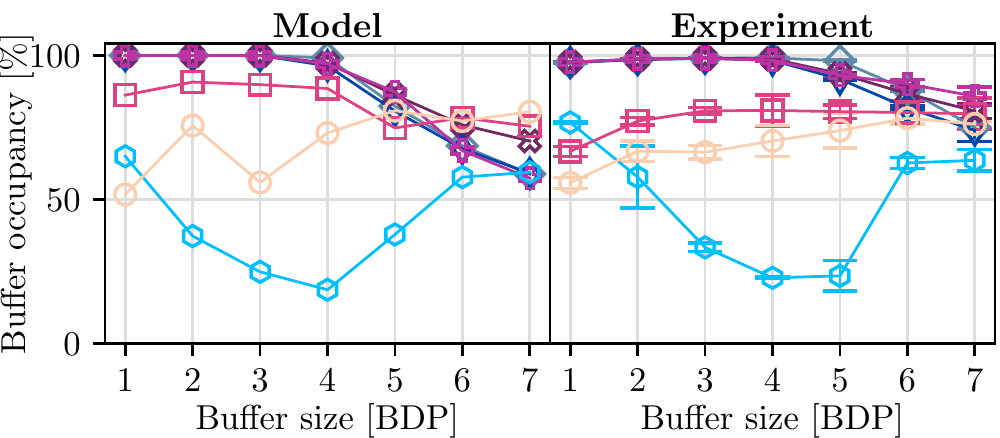}
        \caption{Drop-tail}
        \label{fig:validation:10ms:queuing:droptail}
    \end{subfigure}
    \begin{subfigure}[b]{0.55\linewidth}
        \centering
        \includegraphics[width=\linewidth]{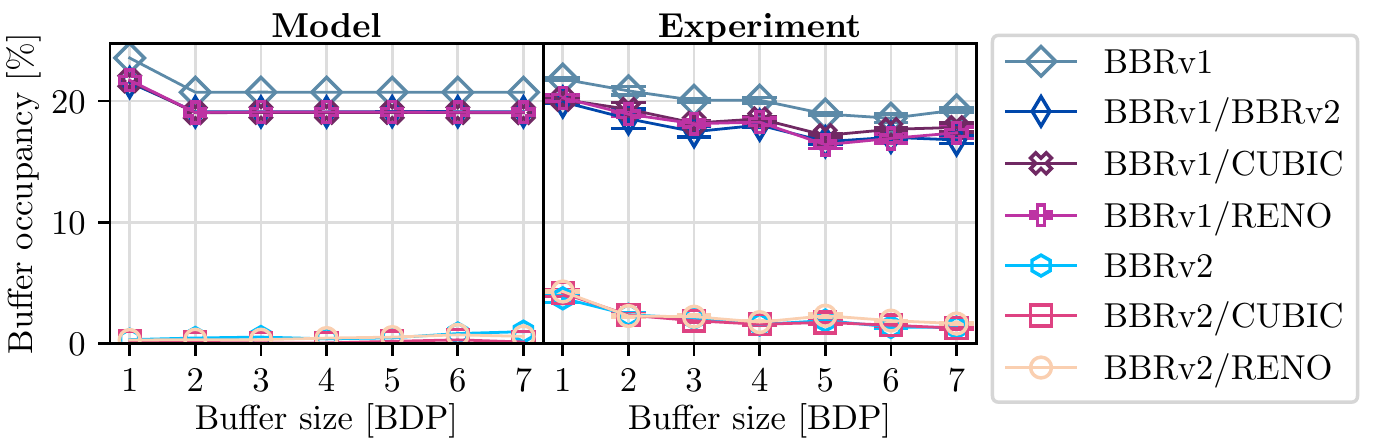}
        \caption{RED}
        \label{fig:validation:10ms:queuing:red}
    \end{subfigure}
    \vspace{-15pt}
    \caption{Queuing validation}
    \label{fig:validation:10ms:queuing}
\end{figure*}

\cref{fig:validation:10ms:queuing} shows the average queue size
as a share of buffer capacity. Notably,
the fluid model captures the effect that the
traditional loss-based CCAs Reno and CUBIC
cause \emph{bufferbloat}, i.e., lead to consistently
high buffer utilization.
Also combined settings of BBRv2 with
loss-based CCAs include high 
buffer utilization, where the generally lower
buffer usage of BBRv2 in a homogeneous setting demonstrates 
that BBRv2 is not responsible for the bufferbloat.

Interestingly, BBRv1 leads to even more intense
buffer usage than loss-sensitive CCAs, whether in
homogeneous settings or in heterogeneous settings
with loss-sensitive CCAs. Under drop-tail 
(\cref{fig:validation:10ms:queuing:droptail}),
BBRv1 uses most of the buffer independent of buffer size,
where the relative buffer usage is only moderately
reduced in large buffers. This effect is surprising,
as a major design goal of BBR is exactly to
avoid the bufferbloat caused by traditional loss-based
CCAs~\cite{cardwell2016bbr}.

The validation analysis reveals another unexpected
phenomenon, which concerns the buffer utilization of BBRv2
in homogeneous settings given a drop-tail queuing discipline.
In particular, BBRv2 leads to constant absolute buffer usage
for buffer sizes up to 4 BDP, which is visible as decreasing
relative buffer usage in~\cref{fig:validation:10ms:queuing:droptail}.
In these scenarios, the adjustments to BBR appear
to have resolved the issue of bufferbloat in BBRv1.
In large buffers, however, the buffer utilization
increases again with buffer size. Through trace
inspection, we found that this phenomenon is caused
by initial measurements of \texttt{inflight\_hi}
during the start-up phase of BBRv2: Given large
buffers, this initial \texttt{inflight\_hi} bound
may be set too high or not set at all. 
Moreover, \texttt{inflight\_lo}
may never be set either, because large buffers prevent
loss, which would activate \texttt{inflight\_lo}. 
In absence of stringent bounds given
by \texttt{inflight\_hi} and \texttt{inflight\_lo},
BBRv2 falls back on the standard BBR congestion-window 
size of 2 estimated BDP (cf.~\cref{eq:bbr2:w_pbw}).
In comparison with the empirically found \texttt{inflight\_hi}
and \texttt{inflight\_lo}, this congestion-window size 
is a loose bound 
that allows higher sending rates of BBRv2 and
thus causes more intense buffering. To the best of
our knowledge, this behavior of BBRv2 has not been
publicly documented so far. While our BBRv2 fluid model
does not model the start-up phase which causes this issue,
the same effect can be observed in the model 
when choosing the initial 
condition of the differential equation for~$w^{\mathrm{hi}}_i$
(cf.~\cref{eq:bbr2:w_hi}) dependent on the buffer size.
Therefore, we note that fluid models have to be evaluated
under a variety of initial conditions to reveal
design issues.

\subsubsection{Utilization}
\label{sec:experiments:aggregate:utilization}

\begin{figure*}
    %\vspace{-10pt}
    \begin{subfigure}[b]{0.4\linewidth}
        \centering
        \includegraphics[width=\linewidth]{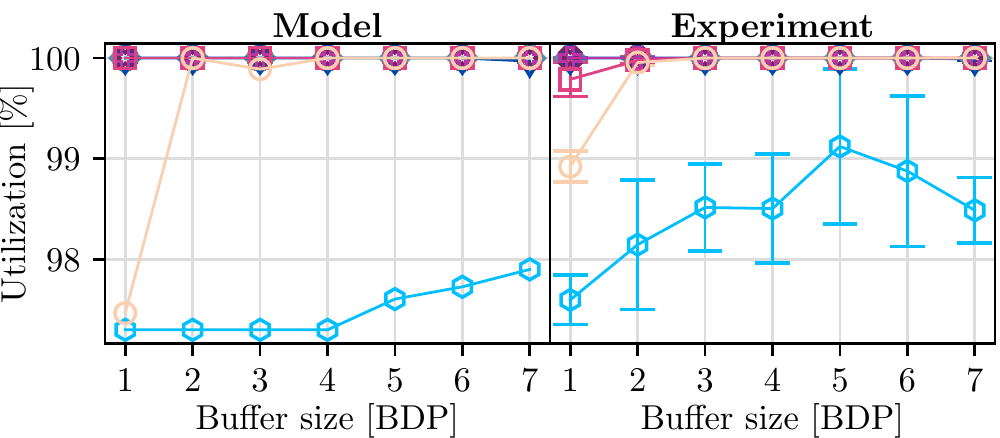}
        \caption{Drop-tail}
        \label{fig:validation:10ms:utilization:droptail}
    \end{subfigure}
    \begin{subfigure}[b]{0.55\linewidth}
        \centering
        \includegraphics[width=\linewidth]{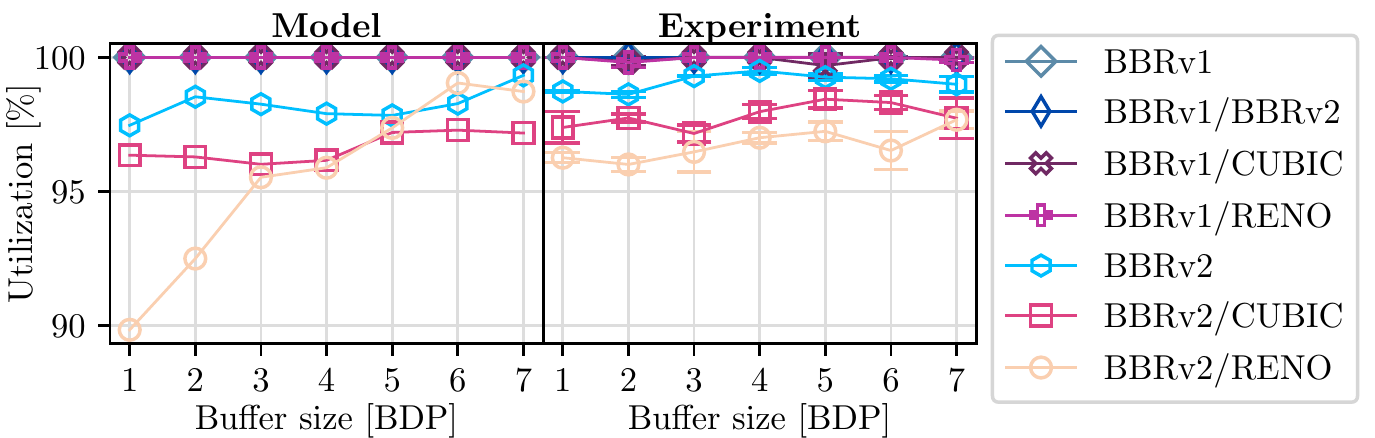}
        \caption{RED}
        \label{fig:validation:10ms:utilization:red}
    \end{subfigure}
    \vspace{-15pt}
    \caption{Utilization validation}
    \label{fig:validation:10ms:utilization}
    \vspace{-5pt}
\end{figure*}

\begin{figure*}
    %\vspace{-10pt}
    \begin{subfigure}[b]{0.4\linewidth}
        \centering
        \includegraphics[width=\linewidth]{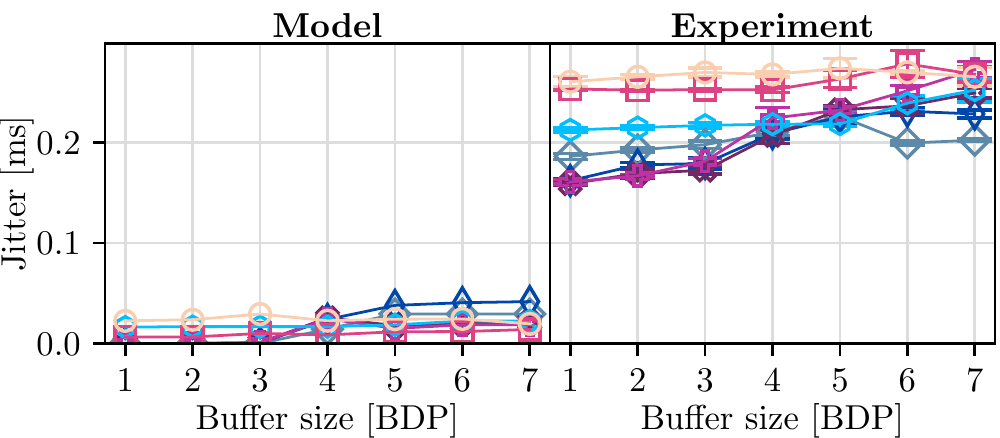}
        \caption{Drop-tail}
        \label{fig:validation:10ms:jitter:droptail}
    \end{subfigure}
    \begin{subfigure}[b]{0.55\linewidth}
        \centering
        \includegraphics[width=\linewidth]{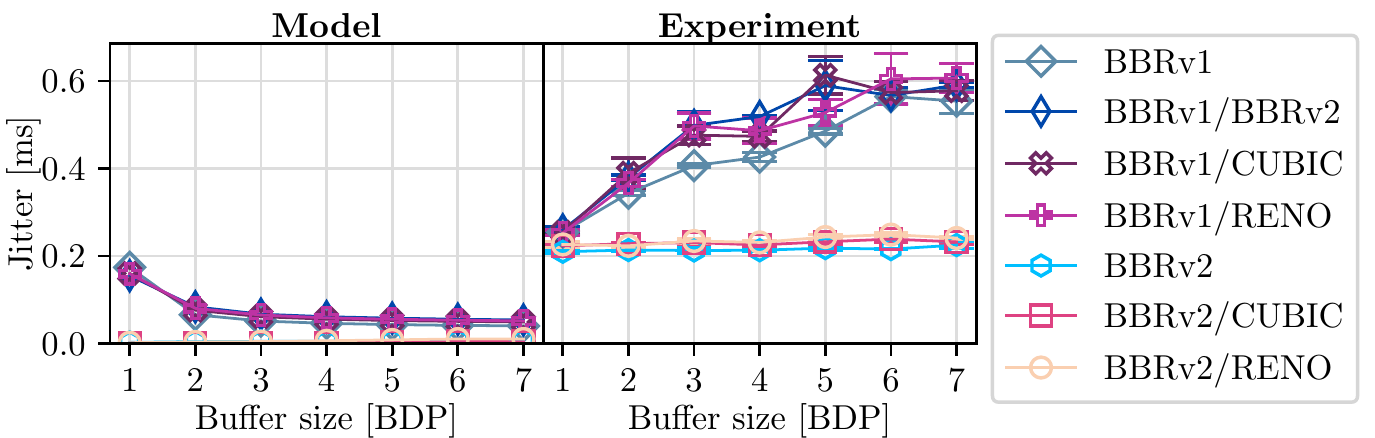}
        \caption{RED}
        \label{fig:validation:10ms:jitter:red}
    \end{subfigure}
    \vspace{-15pt}
    \caption{Jitter validation}
    \label{fig:validation:10ms:jitter}
    \vspace{-5pt}
\end{figure*}

% While fairness, loss and queuing delay are metrics that
% are primarily important from a user perspective,
% the link utilization achieved by CCAs is a key metric
% for network operators. 
The fluid model captures three important aspects
of link utilization~(cf. \cref{fig:validation:10ms:utilization}).
First, the fluid model correctly predicts that BBRv1 
(or combinations with BBRv1) lead to full utilization of
the bottleneck link, both under drop-tail and RED. 
This high utilization by BBR
is unsurprising given the aggressiveness of the CCA,
which also manifests in high loss (\cref{fig:validation:10ms:loss})
and intense queuing~(\cref{fig:validation:10ms:queuing}).
Second, the fluid model mirrors the increasing link utilization
by loss-sensitive CCAs for increasing buffer size under drop-tail: 
Loss-based CCAs grow their rate while the buffer is filling up
and cut it by a constant factor when the buffer is full,
so larger buffers imply higher rates and thus higher utilization.
Third, the fluid model reflects that BBRv2 yields
the lowest utilization given drop-tail among all CCAs, 
although the wasted capacity only amounts to 3\% at most. 
This incomplete utilization stems from the \texttt{ProbeRTT} 
state of BBRv2, in which the inflight is reduced 
to half the estimated BDP. Under RED, BBRv2 does
not enter the \texttt{ProbeRTT} phase for 10 senders;
neither does BBRv1 for both queueing disciplines.

The only major prediction error of the model
is the underestimation
of utilization by the BBRv2/Reno combination
in shallow RED buffers. This result
indicates an exaggeration of the loss
reaction in the Reno fluid model by 
Low et al.~\cite{low2002internet}.

\subsubsection{Jitter}
\label{sec:experiments:aggregate:jitter}

Jitter corresponds to the mean
delay difference between consecutive
packets. The experimental jitter results
in~\cref{fig:validation:10ms:jitter}
are calculated in this manner.
As the fluid model misses a notion of
packets, we compute the jitter for fluid-model traces by
sampling the RTT at a virtual packet rate,
i.e., every $g\cdot N/C_{\ell}$ seconds,
where~$g$ is a given packet size.

However, as~\cref{fig:validation:10ms:jitter}
makes clear, also this makeshift
calculation unsurprisingly fails to predict jitter:
Fluid models intentionally abstract from 
small-scale fluctuations and 
describe only the macroscopic tendency of
network indicators with smooth curves.
% Put differently, fluid models eliminate
% jitter \emph{by design}. 
Nonetheless, fluid models could be combined
with packet-level models aimed at
modeling 
jitter~\cite{dahmouni2012analytical,daniel2003interarrival};
we leave this challenge for future research.

\section{Theoretical Analysis}
\label{sec:analysis}

In this section, we analyze the BBR fluid models to characterize the stability
of these CCAs. The proofs of all
theorems are presented in~\cref{sec:analysis-proofs}.

\subsection{BBRv1 Stability Analysis}
\label{sec:analysis:bbr1}

While the fluid model in~\cref{sec:model:bbr:bbrv1} is suitable
for simulation, we have to simplify it to obtain a high-level model suitable
for analysis (\cref{sec:analysis:bbr1:reduction}).
In~\cref{sec:analysis:bbr1:stability}, we investigate
the existence, form and stability of BBRv1 equilibria.

\subsubsection{Model Reduction}
\label{sec:analysis:bbr1:reduction}

The first simplification step is given by disregarding the \texttt{ProbeRTT} state,
which lets agent discover their propagation delay and generally achieves this goal. 
Hence, we assume that $\tau_i^{\min} = \sum_{\ell\in\pi_i} d_{\ell} =:  d_i$.
Since this minimum-RTT measurement is present 
in the congestion-window size~$w_i^{\mathrm{pbw}} = 2\tau_i^{\min}x_i^{\mathrm{btl}}$,
also~$w_i^{\mathrm{pbw}}$ can be simplified to~$2 d_i x_i^{\mathrm{btl}}$.
Apart from affecting the congestion-window size, the \texttt{ProbeRTT} state
has no lasting effects and can hence be omitted 
for the purpose of stability analysis.

The second step involves understanding the evolution of the maximum
measurement~$x_i^{\max}$. This maximum measurement is the
maximum delivery rate~$x_i^{\mathrm{dlv}}$ over a period, which in turn 
depends on the sending rates of all agents and 
the queue length at the bottleneck link~(cf.~\cref{eq:bbr:x_dlv}).
The sending rate of an agent~$i$ is 
either~$\min(w_i^{\mathrm{pbw}}/\tau_i,\ \nicefrac{5}{4} x_i^{\mathrm{btl}})$ 
(if agent~$i$ is probing),
$\min(w_i^{\mathrm{pbw}}/\tau_i,$ $\nicefrac{3}{4} x_i^{\mathrm{btl}})$ 
(if draining),
or $\min(w_i^{\mathrm{pbw}}/\tau_i,\ x_i^{\mathrm{btl}})$ (otherwise).
Hence, the maximum delivery rate depends on
the concurrent behavior of the other agents on the bottleneck link, 
which may be probing or draining during the measurements of agent~$i$. 
For many agents,
the probing and draining agents can be expected to offset each other
such that the total background-traffic volume
is similar as if~$x_j^{\mathrm{pbw}} = \min(w_j^{\mathrm{pbw}}/\tau_j,\ x_j^{\mathrm{btl}})$
for all agents~$j \in U_{\ell_i}, j \neq i$.
Given this background traffic, an agent~$i$ 
measures the maximum delivery rate in the probing phase.
Hence, the maximum measurement is determined as follows:
\begin{equation}
    x_i^{\max} = \begin{cases}
        \frac{\min(\nicefrac{5}{4},\ \Delta_i)\cdot x_i^{\mathrm{btl}} \cdot C_{\ell_i}}{\min(\nicefrac{5}{4},\ \Delta_i) \cdot x_i^{\mathrm{btl}} + \sum_{j\neq i} \min(1,\ \Delta_j)\cdot x_j^{\mathrm{btl}}} & \text{if } q_{\ell_i} > 0\\
        \min(\nicefrac{5}{4}, \Delta_i)\cdot x_i^{\mathrm{btl}} & \text{otherwise}
    \end{cases}
    \label{eq:analysis:bbr1:xmax}
\end{equation} where~$\Delta_i = 2 d_i / (d_i + \sum_{\ell\in\pi_i} q_{\ell}/C_{\ell})$,
and $\ell_i$ is the bottleneck link for agent~$i$.

The third and final simplification step concerns the adaptation of the bottleneck-bandwidth
estimate~$x_i^{\mathrm{btl}}$. Considered over a long duration, this regular update
can be approximated by a continuous assimilation:
\begin{equation}
    \dot{x}_i^{\mathrm{btl}} = x_i^{\mathrm{max}} - x_i^{\mathrm{btl}}
    \label{eq:analysis:bbr1:xbtl-adaptation}
\end{equation}

\subsubsection{Stability Analysis}
\label{sec:analysis:bbr1:stability}

To characterize stability of a CCA, we first need to identify its \emph{equilibria},
i.e., configurations from which the fluid-model dynamics cannot depart.
In the case of BBRv1, the network state may change if 
the maximum measurement~$x_i^{\max}$ by some agent~$i$ differs from the bottleneck-bandwidth
estimate~$x_i^{\mathrm{btl}}$, or 
if the queue length~$q_\ell$ of some link~$\ell$ grows; both events may lead to
subsequent rate changes. To formalize this condition, we henceforth
consider~$N$ senders which share a single bottleneck link~$\ell^{\ast}$.
Moreover, we first assume that buffer capacities do not constrain the dynamics, 
and modify this assumption later.

\begin{definition}
    $N$ BBRv1 senders sharing a bottleneck link~$\ell^{\ast}$ are in equilibrium
    if and only if $\{x_i^{\mathrm{btl}}\}_{i\in U_{\ell^{\ast}}}$ and~$q_{\ell^\ast}$ satisfy:
    \begin{equation}
        \sum_{i\in U_{\ell^{\ast}}} \min(1,\ \Delta_i) \cdot x_i^{\mathrm{btl}} = C_{\ell^\ast} \quad
        \forall i \in U_{\mathrm{\ell^{\ast}}}.\ x_i^{\mathrm{btl}} = x_i^{\max}
        \label{eq:analysis:bbr1:equilibrium}
    \end{equation}
    \label{def:analysis:bbr1:equilibrium}
\end{definition}

The first condition keeps the aggregate rate~$y_{\ell^\ast}$  at line rate~$C_{\ell^\ast}$ 
and hence ensures a static queue length.
The remaining constraints rule out rate adaptations.
% These conditions imply the following equilibria:
\begin{theorem}
    $N$ BBRv1 senders sharing a bottleneck link~$\ell^{\ast}$ are in
    equilibrium if and only if propagation delay equals queuing delay
    for every sender, i.e., 
    $$\forall i\in U_{\mathrm{\ell^\ast}}.\ d_i = \sum_{\ell\in\pi_i} \frac{q_\ell}{C_\ell}.$$
    \label{thm:analysis:bbr1:equilibrium}
\end{theorem}

Interestingly, \cref{thm:analysis:bbr1:equilibrium} suggests that
the equilibria of BBRv1 in single-bottleneck scenarios (with non-limiting buffers) 
can be arbitrarily unfair as long as~$\sum_{i\in U_{\ell^\ast}}x^{\mathrm{btl}}_i = C_{\ell^{\ast}}$.
% In the most extreme case, an equilibrium can consist of one sender~$i$
% sending at link capacity~$C_{\ell^{\ast}}$ and all other senders transmitting
% zero volume. 
Furthermore, we note that the BBRv1 equilibrium requires
equal path propagation delay~$d$ for all senders if all senders only
encounter a non-empty queue at the bottleneck link~$\ell^{\ast}$.
For our stability analysis, we focus on that case, i.e., a scenario 
where the queue lengths on all involved links except the bottleneck link~$\ell^{\ast}$ 
are zero,  which is a scenario frequently investigated in the 
literature~\cite{addanki2022powertcp,hollot2001control,zhu2016ecn}.
In this case, we can prove asymptotic stability of BBRv1
with the indirect Lyapunov method, meanining that
initial configurations exist for which the
BBRv1 dynamics converge to the equilibrium.

\begin{theorem}
    In a single-bottleneck network with a queue exclusively at the bottleneck,
    the BBRv1 equilibrium from~\cref{thm:analysis:bbr1:equilibrium} is
    asymptotically stable.
    \label{thm:analysis:bbr1:stability}
\end{theorem}

As the proof of~\cref{thm:analysis:bbr1:stability} in~\cref{sec:proof:bbr1:stability} 
makes clear, the BBRv1 equilibrium from~\cref{thm:analysis:bbr1:equilibrium} is only viable 
if the bottleneck-link buffer capacity~$B_{\ell^\ast}$ permits
the equilibrium queue length $q_{\ell^\ast} = dC_{\ell^\ast}$.
Intuitively, the equilibrium is valid for a bottleneck buffer
that is large enough for the congestion-window constraint~$\Delta_i$
to have an impact. 
To analytically investigate the shallow-buffer case
where the congestion-window limit is not effective,
we assume that the bottleneck queue length~$q_{\ell^\ast}$
is restricted by the buffer size~$B_{\ell^\ast}$ such 
that the congestion-window limit has no effect for any
agent~$i$, i.e.,~$\Delta_i \geq \nicefrac{5}{4}$ for 
all~$i \in U_{\ell^\ast}$ (cf.~\cref{eq:analysis:bbr1:xmax}).
With this assumption, we find a different equilibrium for BBRv1:

\begin{theorem}
    $N$ BBRv1 senders sharing a bottleneck link~$\ell^\ast$ that has a
    shallow buffer (i.e.~$\Delta_i > \nicefrac{5}{4}$ $\forall i \in U_{\ell^\ast}$) 
    are in equilibrium if and only if each agent~$i$ has the
    following bottleneck-bandwidth estimate~$x_i^{\mathrm{btl}}$:
    $$x_i^{\mathrm{btl}} = \frac{5C_{\ell^\ast}}{4N+1}.$$
    This equilibrium is perfectly fair and asymptotically stable.
    \label{thm:analysis:bbr1:equilibrium:shallow}
\end{theorem}

\cref{thm:analysis:bbr1:equilibrium:shallow} thus implies that without
an effective congestion-window limit, the aggregate rate~$y_{\ell^\ast}$ in
equilibrium consistently exceeds the link capacity~$C_{\ell^\ast}$,
except for~$N = 1$. As a result, multiple BBRv1 senders
fill the shallow bottleneck-link buffer, eventually incurring a loss
rate equal to the excess sending rate (20\% for $N\rightarrow\infty$). 
While this consistent packet loss does
not reduce the rate of loss-insensitive BBRv1 senders, 
the loss is fatal for loss-based
CCAs on the same bottleneck link, which produces high inter-CCA unfairness.
Among each other, BBRv1 flows forcibly converge to perfect fairness
in shallow buffers, where such fairness is only possible, 
but not required
by the BBRv1 equilibria for deep buffers (cf.~\cref{thm:analysis:bbr1:equilibrium}).

\subsection{BBRv2 Stability Analysis}
\label{sec:analysis:bbr2}

Similar to the preceding section, this section
presents a condensed version of the
BBRv2 fluid model from~\cref{sec:model:bbr:bbrv2},
which is then used for stability analysis.

\subsubsection{Model Reduction}
\label{sec:analysis:bbr2:reduction}

Thanks to the shared foundation of BBRv1
and BBRv2, our reduced fluid model for 
BBRv2 largely matches the reduced model for BBRv1 
(cf.~\cref{sec:analysis:bbr1:reduction}) 
such that we only discuss
different simplification steps.

In particular, the specific probing process of BBRv2
affects the maximum 
measurement~$x_i^{\mathrm{max}}$.
This probing process is centered around
a traffic pulse, which raises
the pacing rate 
to~$\nicefrac{5}{4}\cdot x_i^{\mathrm{btl}}$
and the inflight volume 
to~$\nicefrac{5}{4}\cdot\overline{w}_i$, 
expect the loss exceeds~2\%.
Since we limit our analysis to
networks with buffers large enough
to prevent loss, the traffic-pulse rate
is:
\begin{equation}
    x_i^{\mathrm{pls}} = \nicefrac{5}{4}\cdot\min(1,\ \delta_i) \cdot x_i^{\mathrm{btl}},\ \delta_i = \frac{d_i}{d_i + \sum_{\ell\in\pi_i} q_{\ell}/C_\ell}
\end{equation}

In addition to this pulse rate, the background
traffic co-determines the maximum delivery 
rate~$x_i^{\max}$. 
This background traffic can be assumed to consist
of flows in cruising mode, in which any BBRv2 flow
spends the vast majority of its lifetime. 
In cruising mode, a BBRv2 agent~$i$
sets its pacing rate to~$x_i^{\mathrm{btl}}$
and keeps its inflight volume at the minimum
of the estimated BDP~$\overline{w}_i$ and
85\% of the upper inflight
bound~\texttt{inflight\_hi} ($w_i^{\mathrm{hi}}$).
Since we can exclude packet loss, 
$w_i^{\mathrm{hi}}$ corresponds to the maximum
measured inflight from the probing pulse,
which is~$\nicefrac{5}{4}\cdot\overline{w}_i$.
Since the estimated BDP~$\overline{w}_i$ is
consistently smaller 
than~$0.85\cdot\nicefrac{5}{4}\cdot\overline{w}_i$,
the sending rate in cruising
mode is:
\begin{equation}
    x_i = \min(1, \delta_i) \cdot x_i^{\mathrm{btl}} 
\end{equation}

Based on the sending rates of pulses and
the cruising mode, the evolution of the
maximum delivery rate~$x_i^{\max}$ is approximated
as follows (with~$\ell_i$ as the
bottleneck link):
\begin{equation}
     x_i^{\max} = \begin{cases}
        \frac{\nicefrac{5}{4} \cdot \min(1,\ \delta_i)\cdot x_i^{\mathrm{btl}} \cdot C_{\ell_i}}{\nicefrac{5}{4} \cdot \min(1,\ \delta_i)\cdot x_i^{\mathrm{btl}} + \sum_{j\neq i} \min(1,\ \delta_j)\cdot x_j^{\mathrm{btl}}} & \text{if } q_{\ell_i} > 0\\
       \nicefrac{5}{4} \cdot \min(1,\ \delta_i)\cdot x_i^{\mathrm{btl}} & \text{otherwise}
    \end{cases}
\end{equation}

\subsubsection{Stability Analysis}
\label{sec:analysis:bbr2:stability}

For BBRv2, the equilibrium conditions
match the equilibrium conditions for BBRv1 (cf.~\cref{def:analysis:bbr1:equilibrium}),
with~$\Delta_i$ substituted by~$\delta_i = \Delta_i/2$.
However, the modified adaptation rule
for~$x_i^{\mathrm{btl}}$ induces a different
equilibrium for BBRv2:
\begin{theorem}
    $N$ BBRv2 senders sharing a bottleneck link~$\ell^{\ast}$ are in a perfectly fair equilibrium if propagation delay and queuing delay for each agent have the following relation: $$\forall i\in U_{\mathrm{\ell^\ast}}.\ \frac{N-1}{4N+1} \cdot d_i = \sum_{\ell\in\pi_i} \frac{q_\ell}{C_\ell}$$
    \label{thm:analysis:bbr2:equilibrium}
\end{theorem}

Importantly, the above equilibrium 
is not necessarily the only equilibrium
for BBRv2, which may thus induce unfair
equilibria like BBRv1. 
Nevertheless, the above BBRv2 equilibrium
has an inter-dependency with the rate 
\emph{distribution}; no BBRv1 equilibrium
involves such an inherent dependency. 

Similar to the BBRv1 equilibrium, however,
the above equilibrium implies 
equal path propagation delay~$d$ for all senders if all 
senders only encounter a non-empty queue at the 
bottleneck link~$\ell^{\ast}$.
For our stability analysis, we thus
again focus on a scenario 
where the queue lengths on all involved links except 
the bottleneck link~$\ell^{\ast}$ 
are zero:
\begin{theorem}
    In a single-bottleneck network with a queue exclusively at the bottleneck,
    the BBRv2 equilibrium from~\cref{thm:analysis:bbr2:equilibrium} is
    asymptotically stable.
    \label{thm:analysis:bbr2:stability}
\end{theorem}

Importantly, if a network has a queue exclusively
at the bottleneck link, \cref{thm:analysis:bbr2:equilibrium}
implies equilibrium queue 
length $q_{\ell^\ast} = \frac{N-1}{4N+1}dC_{\ell^\ast}$.
In comparison with BBRv1, BBRv2 thus reduces buffer utilization
by at least 75\% (for~$N\rightarrow\infty$),
assuming the buffer is large enough to accommodate
the BBRv1 equilibrium queue.

%%%%%%%%%%%%%%%%%%%%%%
\section{Insights and Discussion}
\label{sec:discussion}

In this section, we summarize
the most interesting insights from our experimental
validation (\cref{sec:experiments}),
% additional simulations (\cref{sec:results}),
and our theoretical analysis (\cref{sec:analysis}).
These insights reflect properties of
CCAs~(\cref{sec:discussion:insights}) and
properties of the fluid-model 
methodology~(\cref{sec:discussions:differences}).

\subsection{Insights into CCA Performance}
\label{sec:discussion:insights}

One of the most consistent findings in the
previous sections relates to the packet
loss caused by different types of CCAs:

\begin{insight}
    \textbf{Loss Rates of CCAs.}
    Both under drop-tail and RED queuing disciplines,
    BBRv1 causes considerable loss of up to 20\%
    of traffic, while the loss-sensitive CCAs Reno, CUBIC,
    and BBRv2 cause loss rates of around 1\%.
\end{insight}

While such a difference between
BBRv1 and loss-based CCAs is not surprising given
different loss sensitivity, the large
extent of the loss caused by BBRv1 is unexpected.
Importantly, while the loss insensitivity
of BBRv1 does not lead to throughput reductions
of BBRv1, the high packet loss will still lead
to unsatisfactory application performance.
Moreover, in competition with loss-sensitive CCAs, 
the loss insensitivity of BBRv1 poses
a fairness concern, which is also reflected
in previous work~\cite{ware2019modeling,scholz2018towards}:

\begin{insight}
    \textbf{BBRv1's Unfairness Towards Loss-Based CCAs.}
    BBRv1 is highly unfair towards loss-sensitive CCAs,
    leading to near starvation of loss-based flows
    in shallow buffers (given a drop-tail queuing discipline)
    or buffers of any size (given a RED queuing discipline).
    In large drop-tail buffers, the congestion window
    of BBRv1 becomes effective, leading to improvements
    in fairness towards loss-sensitive CCAs.
\end{insight}

The aggressiveness of BBRv1, causing
high loss and unfairness, has two other 
effects:

\begin{insight}
    \textbf{Utilization and Buffer Usage of BBRv1.}
    In all investigated settings, BBRv1 (also in
    combination with other CCAs) achieves full
    link utilization, but also significant
    bufferbloat independent of the queueing discipine.
\end{insight}

Most of these insights gained from our fluid model
have already been identified by previous, 
experiment-based analyses. In response to documentations
of these issues, Google has begun to develop BBRv2,
which can be characterized as follows:
\begin{insight}
    \textbf{Performance of BBRv2.} BBRv2 mostly
    achieves the redesign goals of reduced buffer
    usage, avoiding excessive loss, and
    preserving fairness to loss-based
    CCAs.
\end{insight}

However, we have identified two settings in which
BBRv2 does not achieve its design goal:
\begin{insight}
    \textbf{BBRv2 in Large Drop-Tail Buffers.} In drop-tail 
    buffers with a size exceeding five BDP, BBRv2 causes
    higher buffer utilization than for smaller buffers,
    caused by distortions in an initial \texttt{inflight\_hi}
    estimate in the start-up phase.
\end{insight}

\begin{insight}
    \textbf{BBRv2 in RED Buffers of High-Capacity Links.} 
    When competing with loss-based CCAs (Reno and CUBIC)
    on a high-capacity link under a RED queuing discipline,
    BBRv2 is unfair towards the loss-based CCAs. The reason
    for this unfairness is that on high-capacity links,
    the loss sensitivity of loss-based CCAs is markedly
    higher than the loss sensitivity of BBRv2.
\end{insight}

\subsection{Insights into Fluid Models}
\label{sec:discussions:differences}

The preceding sections 
not only yield
valuable insights into the performance
characteristics of BBR, but also
illustrates the strengths and limitations
of fluid models as an analysis tool. We assess the predictive
power of fluid models as follows:

\begin{insight}
    \textbf{Qualitative Accuracy of Fluid Models.} 
    Fluid models are highly predictive from a
    qualitative perspective, i.e.,
    they accurately capture the direction of correlations
    between CCA performance and network parameters
    as well as the ranking of different CCAs according
    to performance metrics.
\end{insight}

\begin{insight}
    \textbf{Quantitative Accuracy of Fluid Models.} 
    The accuracy of the quantitative predictions
    by fluid models depends on the metric: 
    While the quantitative predictions of fluid
    models are highly accurate regarding loss and
    fairly accurate regarding buffer usage,
    the quantitative predictions regarding fairness
    and utilization are only partially accurate.
\end{insight}

Despite their overall high predictive power,
fluid models yield misleading results in some
cases. We identified the following sources of 
potentially inaccurate predictions:
\begin{insight}
    \textbf{Sources of Inaccuracy.}
    Inaccurate predictions by fluid models can result
    from at least three sources:
    \begin{itemize}
        \item idealizations, e.g., assuming instantly reacting RED queues
        (cf.~\cref{sec:experiments:aggregate:loss});
        \item difficulty of capturing discrete phenomena, e.g.,
        jitter (cf.~\cref{sec:experiments:aggregate:jitter}); and
        \item negligence of the start-up phase, e.g., BBRv2
        has to be simulated with varying initial conditions
        to find issues arising from the start-up phase
        (cf.~\cref{sec:experiments:aggregate:queueing}).
    \end{itemize}
\end{insight}

If the developers of CCAs are aware of the above
pitfalls of fluid models, they can interpret
the fluid-model results in the context
of these caveats.
%%%%%%%%%%%%%%%%%%%%%%
\section{Related Work}
\label{sec:related}

% Congestion is one of the most fundamental challenges in networks and has been studied intensively in the literature. 
While our focus on this paper is on congestion-control algorithms (CCAs), we note that there exist several orthogonal approaches to deal with congestion, for example, related to buffer management~\cite{choudhury1998dynamic,apostolaki2019fab,cisco9000}, scheduling~\cite{alizadeh2013pfabric,hong2012finishing,perry2014fastpass,perry2017flowtune},
and bandwidth reservation~\cite{brown2020future,basescu2015sibra}.

Since the seminal work by Jacobson~\cite{jacobson1988congestion}, a wide range of CCAs have been proposed and analyzed~\cite{hollot2001control,kelly1998rate,huang2006generalized,poojary2019asymptotic,bao2010model,jin2004fast}. While traditional CCAs are based on loss (timeout) signals, more recent protocols
% , especially in datacenters\footnote{CC is considered ``a key enabler (or limiter) of system performance in datacenters''~\cite{kumar2020swift}.}---e.g., DCTCP, D$^2$TCP---
leverage explicit congestion notification (ECN) \cite{alizadeh2010data,zhu2015congestion,vamanan2012deadline} or delay~\cite{mittal2015timely,kumar2020swift,hayes2011revisiting,lee2015accurate,addanki2022powertcp} 
 to react in a more informed and fine-grained manner.
% Also delay has become a popular congestion signal, with CC protocols exploiting the availability of accurate packet timestamps in NICs, and measuring  round-trip times to update congestion windows~\cite{mittal2015timely,kumar2020swift,hayes2011revisiting,lee2015accurate}.
% Protocols such as XCP~\cite{katabi2002congestion}, D$^3$~\cite{wilson2011better}, and RCP~\cite{dukkipati2006flow} rely on explicit network feedback based on rate calculations within the network. In order to improve the fidelity of network feedback, in-band telemetry can be used successfully~\cite{li2019hpcc}.
% A fast reacting CC has recently also become important in the context of emerging reconfigurable networks~\cite{mukerjee2020adapting}.
With BBR~\cite{cardwell2016bbr}, recently another flavor of CC has been introduced, which is often referred to as model-based.

% Olsen categorized existing TCP models into into (i) renewal-theory models, (ii) fixed-point methods, (iii) fluid models, (iv) pro\-ces\-sor-sharing models, and (v) control-theoretic models~\cite{olsen2003stochastic}.
Fluid models (also known as differential-equation models) provide a particularly powerful framework for an analytical understanding of CC protocols and their equilibria~\cite{srikant2004mathematics}, and have been widely used in the literature~\cite{liu2003fluid,low2002internet,vardoyan2018towards,misra2000fluid,foreest2003analysis,raina2005buffer}.
These models are attractive for their flexibility (e.g., supporting different topologies and queuing disciplines), and for allowing fast initial analyses.
In general, the models come in different flavors and can for example be analyzed using
dynamical-systems techniques \cite{raina2006congestion}.
In one prominent work~\cite{misra2000fluid}, a dynamic model of TCP behavior is proposed
using a fluid-flow and stochastic differential-equation analysis.
Using the Runge-Kutta algorithm, the fluid model also allows efficient time-stepped network simulations~\cite{liu2003fluid}.
% We note that our model simulation goes far beyond existing evaluations
% in the literature on fluid models, partially enabled by
% the growth of computational resources since the
% time when fluid models were a research focus.
However, we are not aware of any work devising fluid models for BBR.

That said, BBR has been studied in a number of papers.
In particular, Hock et al.~\cite{hock2017experimental} present a first independent study of BBRv1
and found fairness issues, and
that multiple BBR flows operate at their in-flight cap in buffer-bloated networks.
This work led to several interesting follow-up works~\cite{dong2018pcc,scholz2018towards,turkovic2019fifty}.
In particular, Scholz et al.~\cite{scholz2018towards} show that BBRv1 flows are robustly able to claim a disproportionate share of the bandwidth.
Ware et al.~\cite{ware2019modeling} recently complement these
empirical studies by presenting a first analytical model (although not based on differential
equations) capturing BBR’s behavior in competition with loss-based CCAs in deep buffers.
Yang et al.~\cite{yang2019adaptive} devise a simple fluid model for Adaptive-BBR, i.e.,
their adaptation of BBR specialized for wireless links. 
Neither of these model-based works possesses the generality of our fluid model, 
nor do they include a rigorous convergence analysis.
BBRv2 has been investigated by a number of experiment-based 
studies~\cite{gomez2020performance,kfoury2020emulation,nandagiri2020bbrvl,song2021understanding},
finding mostly that BBRv2 resolves the most serious issues of BBRv1, but also
identifying problematic facets of BBRv2 behavior, although not the ones
found by this paper.

%Google is actively developing BBRv2 and very recently released
%a Linux kernel implementation of BBRv2 [2, 7, 8]. Early presentations
%[8] imply that it primarily resolves the fairness issues discussed
%by Hock et al [11], but does not touch on the fixed proportion
%of link capacity as discussed in this paper.

Recently, CCA research methodology
has experienced innovation with
promising proposals
for an axiomatic approach~\cite{zarchy2019axiomatizing}
and a formal-verification approach~\cite{arun2021formally}.
These approaches are complementary
to the fluid-model approach: 
While the axiomatic
approach allows to identify
fundamental design constraints
and the formal-verification approach
allows to identify network configurations
in which CCA performance does
not conform to specifications,
neither of them is equally well-suited
as the fluid-model approach
to reveal the qualitative and
quantitative effects of
network settings and competing
CCAs on CCA performance.

%%%%%%%%%%%%%%%%%%%%%%
\section{Conclusion}
\label{sec:conclusion}

In this paper, we take a deep dive into the
recent CCA proposals of BBRv1 and BBRv2
by complementing previous analyses with
an approach based on \emph{fluid models}.
Fluid models are a classic but lately seldom employed
approach to evaluating CCA properties,
and are unique in their ability to allow both
theoretical stability analysis and
efficient simulation for a wide range of network scenarios.
We devise such a fluid model for both BBR versions
by using new modelling techniques such as sigmoid pulses
and mode variables, and perform an experiment-based
validation to show that the model is highly predictive
regarding performance and fairness properties.
We further leverage the model for both an extensive
simulation and a theoretical stability analysis.
This investigation confirms previously found issues in BBRv1,
but also yields new insights, e.g., regarding the structure
and asymptotic stability of BBR equilibria, as well as 
regarding bufferbloat and inter-CCA unfairness in BBRv2.

While our model is accurate and general, 
we understand our analysis as a first step in exploring
the investigation opportunities that our fluid model 
opens up. Indeed, it will be interesting to
evaluate the BBR fluid models in multiple-bottleneck
scenarios, both through simulations and further theoretical
analysis. To facilitate such follow-up work, 
we will make all our 
software available as open-source 
together with the final version of the paper.

\bibliographystyle{ACM-Reference-Format}
\bibliography{ref}

\appendix
\clearpage

\section{Ethics}
This work raises no ethical issues.

\section{Loss-Based CCA Models}
\label{sec:model-cc}

In this section, we discuss the existing CCA models for Reno and CUBIC,
which have been used in our simulations. We also validate these
CCA fluid models in isolation

\subsection{Reno}
\label{sec:model-reno}
In its congestion-avoidance phase,
TCP Reno increases the congestion-window size
by $1/w$ upon successful
transmission (signaled by an ACK) and cuts it
in half upon loss. This adaptation
logic is approximated by the following differential equation
for the congestion-window size~\cwnd{t} of agent~$i$
using path~\pathi{}~\cite{low2002internet}:
\begin{equation}
\begin{split}
    \dot{w_i} &= \rate{t-\propdp{}}\cdot \left(1 - \loss[i]{t-\propdp{}}\right) \cdot \frac{1}{\cwnd{}}\\
    %\\ &\hphantom{{}={}}
    & - \rate{t-\propdp{}} \cdot \loss[\pathi]{t-\propdp{}} \cdot \frac{w_{i}}{2}
    %&=\rate{t-\propdp{}}\cdot\left[\frac{1}{\cwnd{t}} - \loss[i]{t-\propdp{}}\cdot\frac{2+\cwnd{t}}{2\,\cwnd{t}^2}\right].
    \label{eq:w:reno}
\end{split}
\end{equation}

\subsection{CUBIC}
\label{sec:model-cubic}

In contrast, TCP CUBIC cannot directly be described
with a differential equation for the congestion-window
size. Instead, Vardoyan et al.~\cite{vardoyan2018towards}
suggest to track
two instrumental variables in CUBIC,
namely the time since last loss of
agent~$i$, $s_i$, and the congestion-window size
at that moment of loss, $w_i^{\max}$:
\begin{subequations}
\begin{align}
    \dot{s}_i &= 1 - s_i \cdot \rate{t-\propdp{}} \cdot \loss[i]{t-\propdp{}},
    \label{eq:s}\\
    \dot{w}_i^{\max} &= \left(\cwnd{} - w_i^{\max}\right) \cdot \rate{t-\propdp{}} \cdot \loss[i]{t-\propdp{}}.
    \label{eq:wmax}
\end{align}
\end{subequations}
The intuition behind \cref{eq:s} is that~$s_i$ is increased by 1
in absence of loss ($p_{\pathi} = 0$) and reduced to 0 when a loss occurs.
\Cref{eq:wmax} describes that~$w_i^{\max}(t)$ should be updated
to~\cwnd{t} in presence of loss. Knowing~$s_i$ and~$w_i^{\max}$,
the congestion-window size~$w$ can be determined by the CUBIC
window-growth function~\cite{ha2008cubic},
\begin{equation}
    \cwnd{} = c\cdot\left(s_i - \sqrt[3]{\frac{w_i^{\max}\cdot b}{c}}\right)^3 + w_i^{\max},
    \label{eq:w:cubic}
\end{equation} where~$c$ and~$b$ are configurable
parameters with standardized values of~$0.4$ and~$0.7$,
respectively~\cite{rfc8312}.
Moreover, the CUBIC implementation
in the Linux kernel uses a time unit
of around 1 second for~$s_i(t)$~\cite{ha2008cubicimplementation}.

\subsection{Trace Validation of Models}

\begin{figure*}
    \begin{subfigure}[b]{0.4\linewidth}
        \centering
        \includegraphics[width=\linewidth]{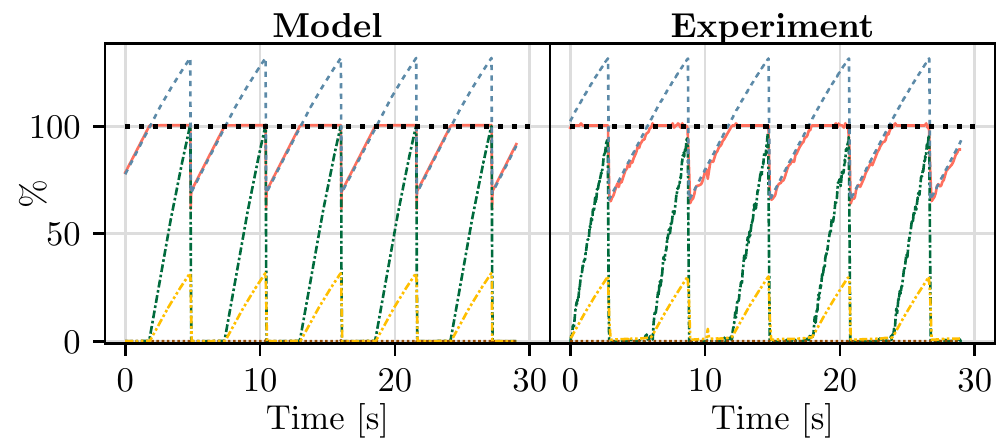}
        \vspace{-15pt}
        \caption{Drop-tail}
        \label{fig:validation:traces:reno:droptail}
    \end{subfigure}
    \begin{subfigure}[b]{0.55\linewidth}
        \centering
        \includegraphics[width=\linewidth]{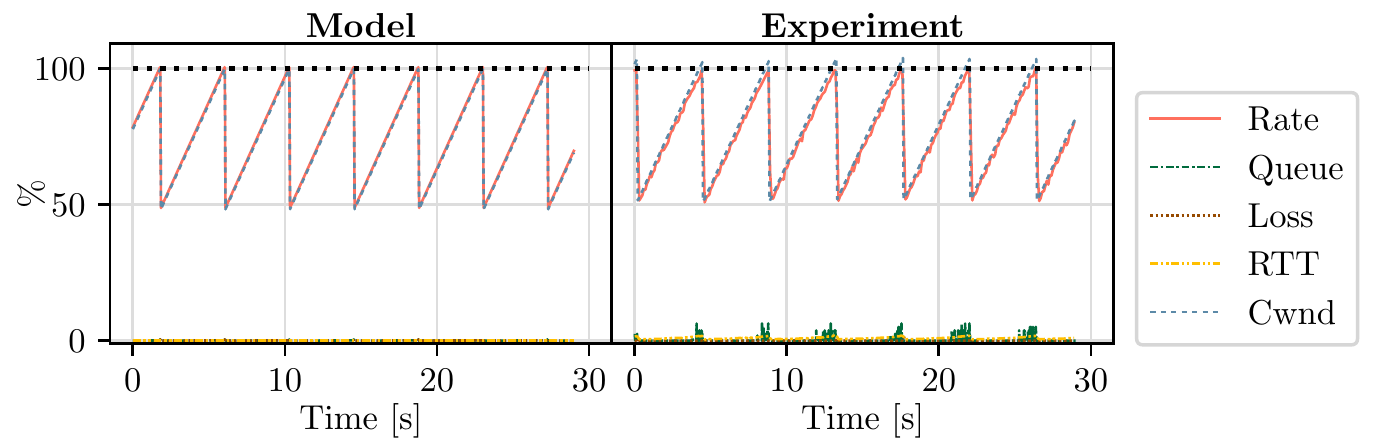}
        \vspace{-15pt}
        \caption{RED}
        \label{fig:validation:traces:reno:red}
    \end{subfigure}
    \vspace{-15pt}
    \caption{Reno trace validation}
    \label{fig:validation:traces:reno}
    \vspace{-5pt}
\end{figure*}

\begin{figure*}
    %\vspace{-10pt}
    \begin{subfigure}[b]{0.4\linewidth}
        \centering
        \includegraphics[width=\linewidth]{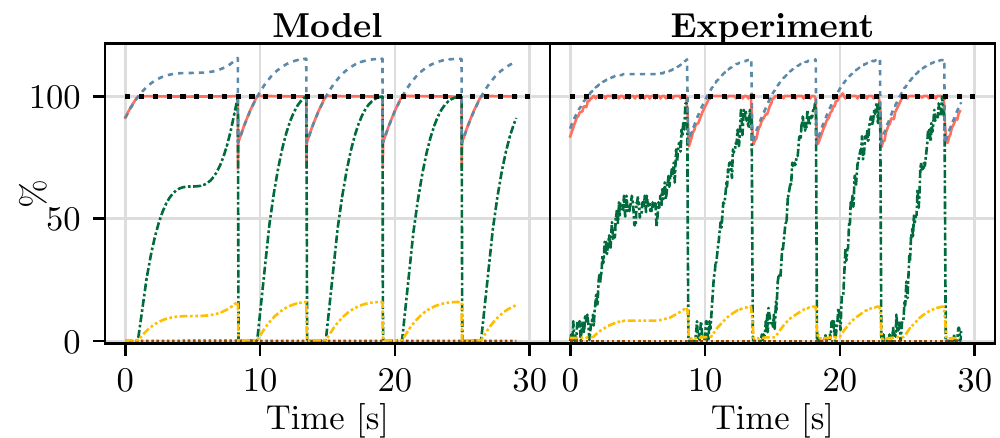}
        \vspace{-15pt}
        \caption{Drop-tail}
        \label{fig:validation:traces:cubic:droptail}
    \end{subfigure}
    \begin{subfigure}[b]{0.55\linewidth}
        \centering
        \includegraphics[width=\linewidth]{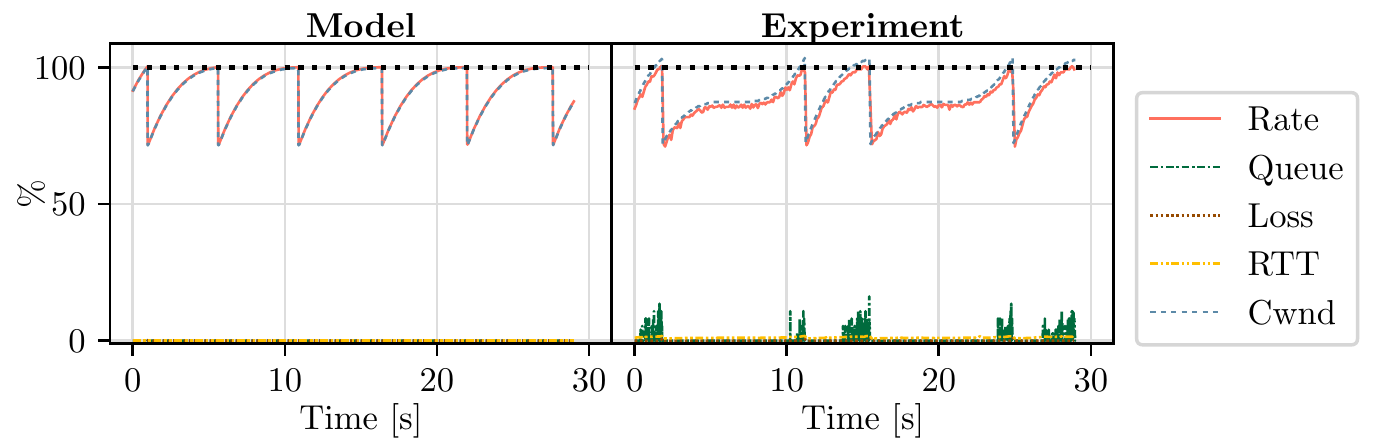}
        \vspace{-15pt}
        \caption{RED}
        \label{fig:validation:traces:cubic:red}
    \end{subfigure}
    \vspace{-15pt}
    \caption{CUBIC trace validation}
    \label{fig:validation:traces:cubic}
    \vspace{-5pt}
\end{figure*}

\cref{fig:validation:traces:reno,fig:validation:traces:cubic}
present a comparison of single-sender traces obtained
from running both model simulation and mininet experiments.
The fluid model correctly
predicts that the rate growth of Reno and 
CUBIC decouples from the congestion-window growth as
soon as the buffer fills up.
In addition, the fluid models correctly 
capture that Reno and CUBIC
lead to considerably smaller loss
(barely visible) than BBRv1, which is
insensitive to loss (cf.~\cref{sec:experiments:traces}).
Finally, the fluid model correctly predicts
that the sending rate of loss-based CCAs never 
exceeds the bottleneck rate under RED, 
while the congestion
windows can temporarily exceed the network
BDP under a drop-tail queuing discipline.
As a result, the smaller
buffer usage under RED is also reflected
in the model, although the difference
between RED and drop-tail is more pronounced
in the model.
This last difference is due to the 
idealization of the RED algorithm in the model.

\section{Aggregate Validation for Short RTT}
\label{sec:aggregate-validation:5ms}

\begin{figure*}[!h]
    \begin{subfigure}[b]{0.4\linewidth}
        \centering
        \includegraphics[width=\linewidth]{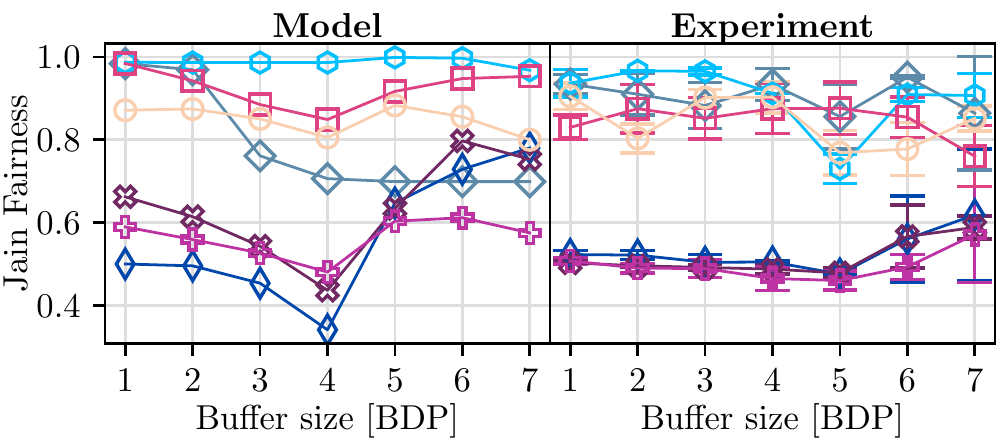}
        \caption{Drop-tail}
        \label{fig:validation:5ms:fairness:droptail}
    \end{subfigure}
    \begin{subfigure}[b]{0.55\linewidth}
        \centering
        \includegraphics[width=\linewidth]{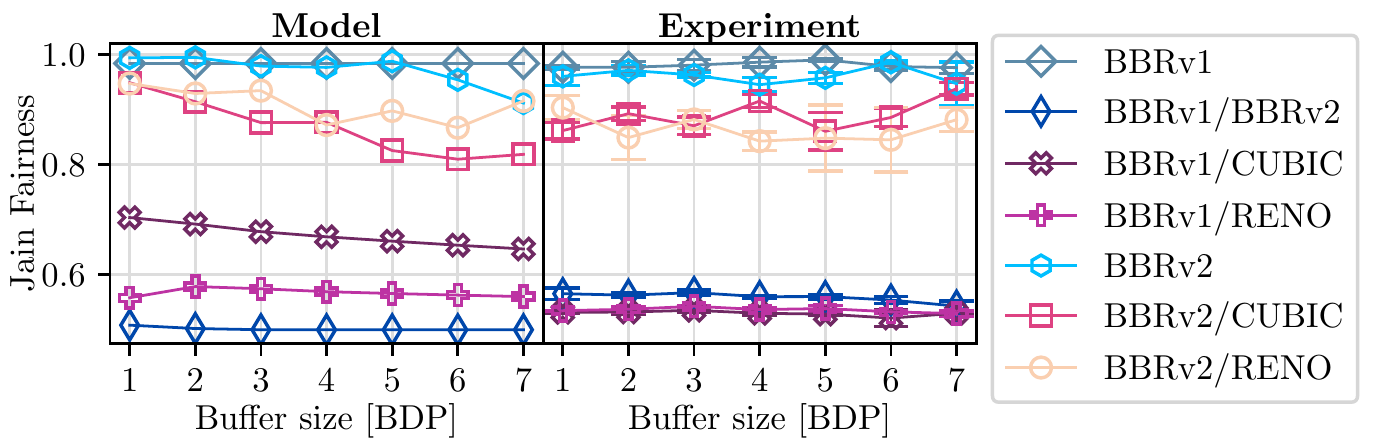}
        \caption{RED}
        \label{fig:validation:5ms:red}
    \end{subfigure}
    \vspace{-15pt}
    \caption{Fairness validation}
    \label{fig:validation:5ms:fairness}
\end{figure*}

\begin{figure*}[!h]
    \begin{subfigure}[b]{0.4\linewidth}
        \centering
        \includegraphics[width=\linewidth]{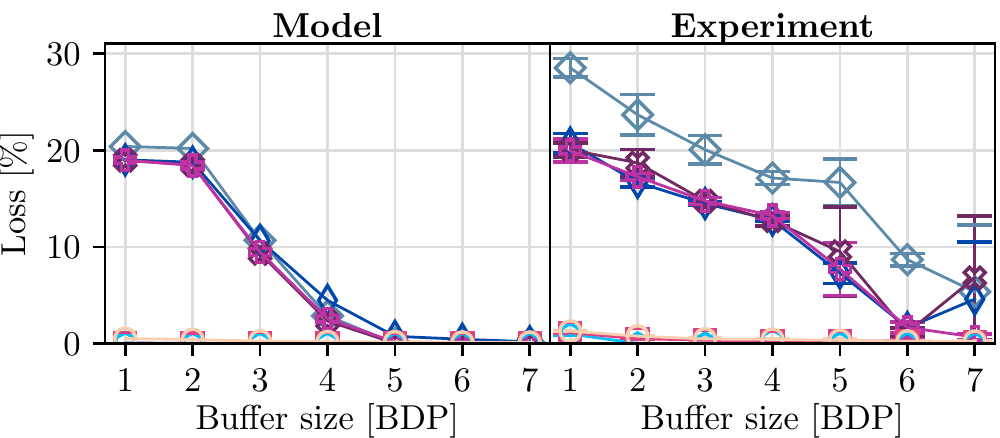}
        \caption{Drop-tail}
        \label{fig:validation:5ms:loss:droptail}
    \end{subfigure}
    \begin{subfigure}[b]{0.55\linewidth}
        \centering
        \includegraphics[width=\linewidth]{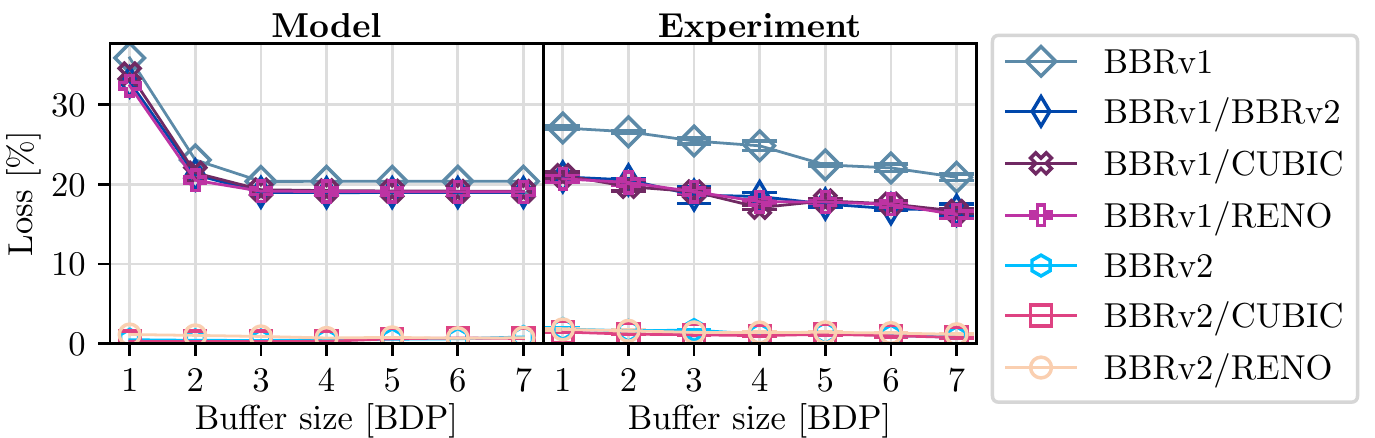}
        \caption{RED}
        \label{fig:validation:5ms:loss:red}
    \end{subfigure}
    \vspace{-15pt}
    \caption{Loss validation}
    \label{fig:validation:5ms:loss}
\end{figure*}

\begin{figure*}[!h]
    %\vspace{-10pt}
    \begin{subfigure}[b]{0.4\linewidth}
        \centering
        \includegraphics[width=\linewidth]{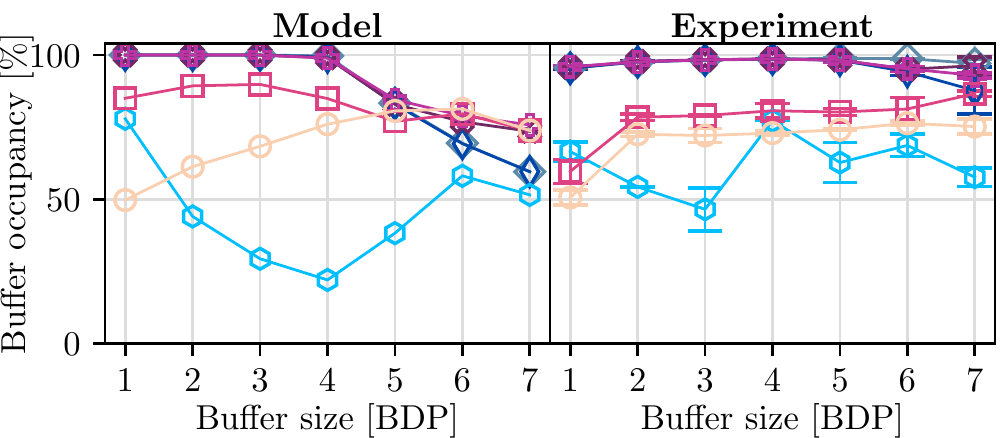}
        \caption{Drop-tail}
        \label{fig:validation:5ms:queuing:droptail}
    \end{subfigure}
    \begin{subfigure}[b]{0.55\linewidth}
        \centering
        \includegraphics[width=\linewidth]{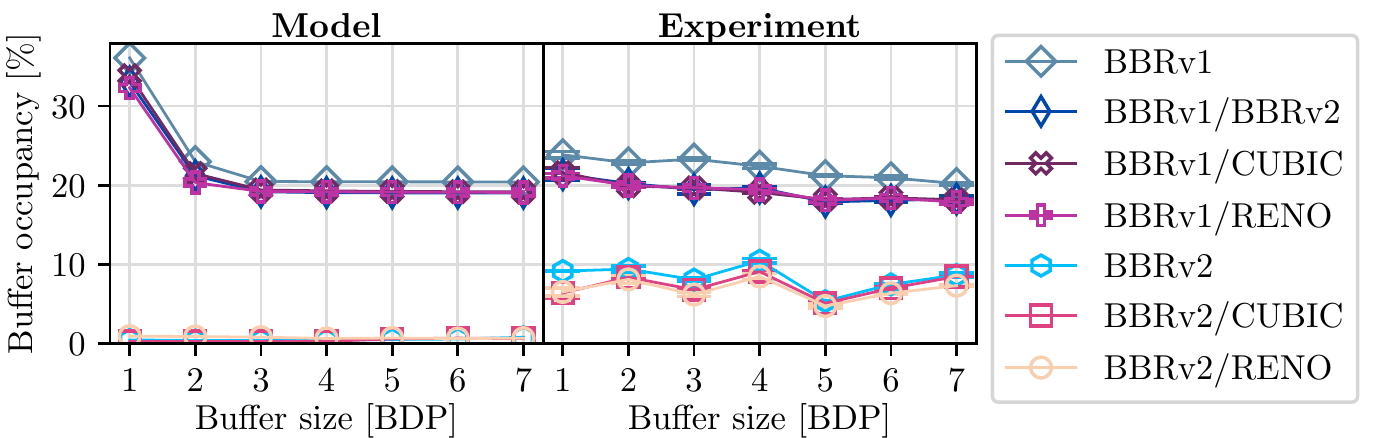}
        \caption{RED}
        \label{fig:validation:5ms:queuing:red}
    \end{subfigure}
    \vspace{-15pt}
    \caption{Queuing validation}
    \label{fig:validation:5ms:queuing}
\end{figure*}

\begin{figure*}[!h]
    %\vspace{-10pt}
    \begin{subfigure}[b]{0.4\linewidth}
        \centering
        \includegraphics[width=\linewidth]{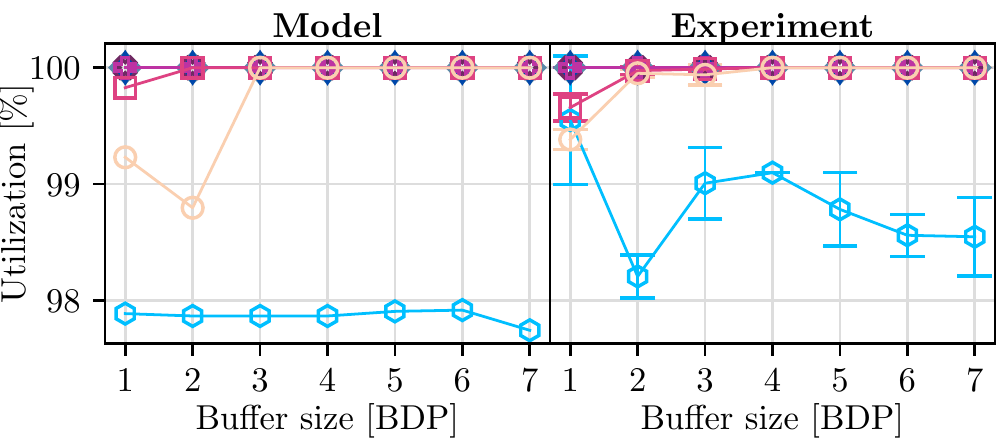}
        \caption{Drop-tail}
        \label{fig:validation:5ms:utilization:droptail}
    \end{subfigure}
    \begin{subfigure}[b]{0.55\linewidth}
        \centering
        \includegraphics[width=\linewidth]{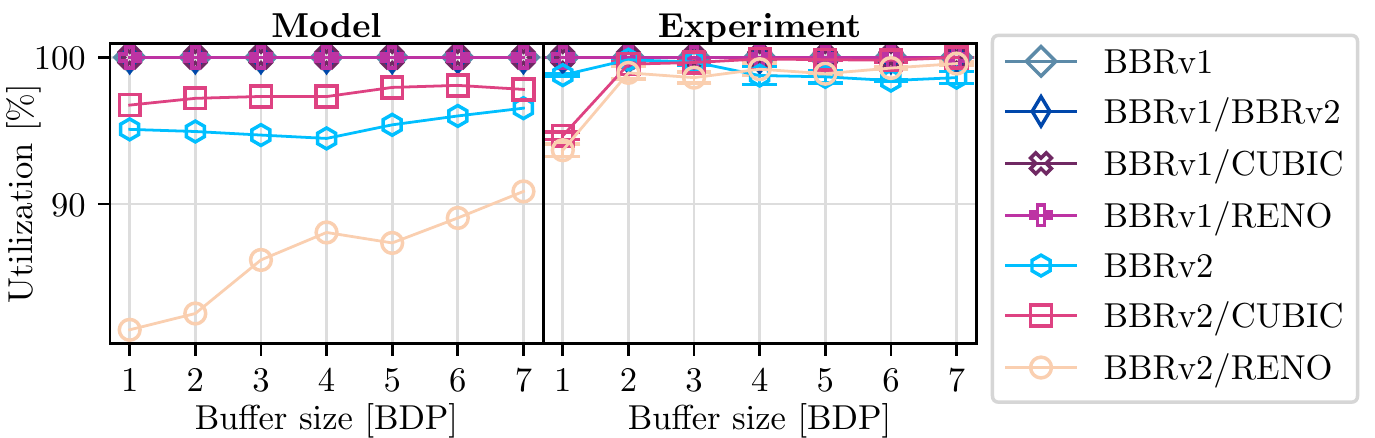}
        \caption{RED}
        \label{fig:validation:5ms:utilization:red}
    \end{subfigure}
    \vspace{-15pt}
    \caption{Utilization validation}
    \label{fig:validation:5ms:utilization}
\end{figure*}

\begin{figure*}[!h]
    %\vspace{-10pt}
    \begin{subfigure}[b]{0.4\linewidth}
        \centering
        \includegraphics[width=\linewidth]{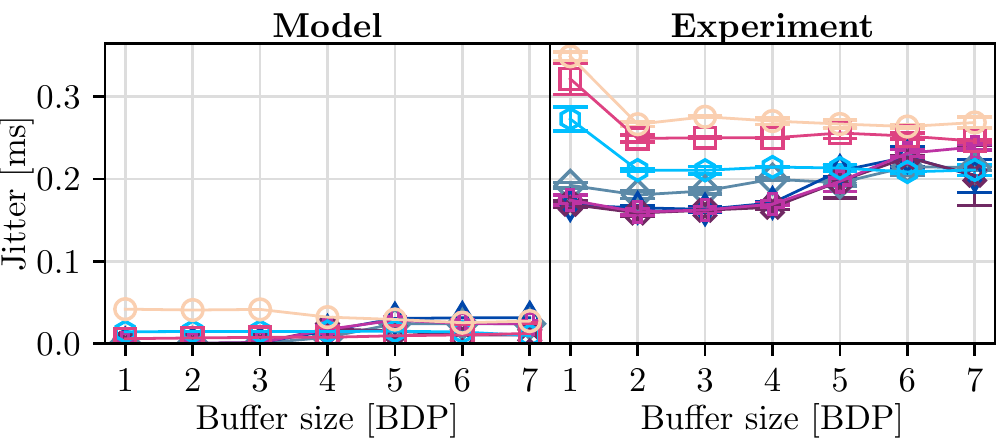}
        \caption{Drop-tail}
        \label{fig:validation:5ms:jitter:droptail}
    \end{subfigure}
    \begin{subfigure}[b]{0.55\linewidth}
        \centering
        \includegraphics[width=\linewidth]{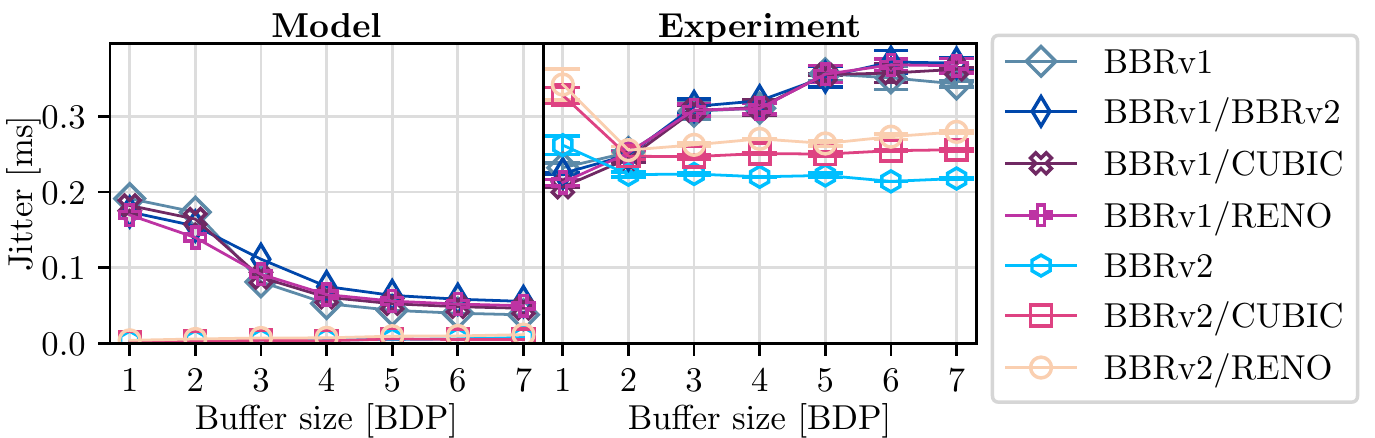}
        \caption{RED}
        \label{fig:validation:5ms:jitter:red}
    \end{subfigure}
    \vspace{-15pt}
    \caption{Jitter validation}
    \label{fig:validation:5ms:jitter}
\end{figure*}

\Cref{fig:validation:5ms:fairness,,fig:validation:5ms:loss,,fig:validation:5ms:queuing,,fig:validation:5ms:utilization,,fig:validation:5ms:jitter} extend the validation, performed
in~\cref{sec:experiments:aggregate}, of fluid models
regarding the aggregate metrics Jain fairness, loss rates,
buffer occupancy, utilization and jitter, respectively.
In contrast to the validation in the body of the paper,
the fluid-model predictions are experimentally
validated for a bottleneck-link delay of 5 milliseconds
and total RTTs between 10 and 20 milliseconds.

\section{Analysis Proofs}
\label{sec:analysis-proofs}

This appendix section contains the proofs of the
theorems in~\cref{sec:analysis}.

\subsection{Proof of~\cref{thm:analysis:bbr1:equilibrium}}
\label{sec:proof:bbr1:equilibria}

First, we consider the case for~$q_{\ell^\ast} = 0$. 
    In that case, the only solution to the conditions in~\cref{eq:analysis:bbr1:equilibrium}
    in terms of~$\{\Delta_i\}_{i\in U_{\ell^{\ast}}}$ is $\Delta_i = 1\ \forall i\in U_{\ell^{\ast}}$,
    which shows that the network is in equilibrium for~$q_{\ell^{\ast}} = 0$ and $d_i = \sum_{\ell\in\pi_i} q_\ell/C_\ell\ \forall i\in U_{\mathrm{\ell^\ast}}$. 
    
    For~$q_{\ell^{\ast}} > 0$, we note that the conditions in~\cref{eq:analysis:bbr1:equilibrium}
    can be transformed into the following conditions for each~$x_i^{\mathrm{btl}}$:
    \begin{align}
         x_i^{\mathrm{btl}} &= \max(1,\ \nicefrac{1}{\Delta_i}) \cdot (C_{\ell^{\ast}} - \sum_{j\neq i} \min(1,\ \Delta_j) \cdot x_j^{\mathrm{btl}}) \label{eq:analysis:bbr1:queue-condition}\\
        x_i^{\mathrm{btl}} &= C_{\ell^{\ast}} - \max(\nicefrac{4}{5},\ \nicefrac{1}{\Delta_i}) \cdot \sum_{j\neq i} \min(1,\ \Delta_j) \cdot x_j^{\mathrm{btl}} \label{eq:analysis:bbr1:rate-condition}
    \end{align}
    Clearly, the previously found solution~$\Delta_i = 1\ \forall i\in U_{\mathrm{\ell^\ast}}$
    is also a solution to the conditions in \cref{eq:analysis:bbr1:rate-condition,eq:analysis:bbr1:queue-condition}.
    Hence, we have proven that the network is in equilibrium if  $d_i = \sum_{\ell\in\pi_i} q_\ell/C_\ell\ \forall i\in U_{\mathrm{\ell^\ast}}$.
    
    It remains to prove that the previously mentioned equilibria are the only possible equilibria.
    To confirm the uniqueness of these equilibria, we first assume an equilibrium 
    where $\exists i\in U_{\mathrm{\ell^{\ast}}}.\ \Delta_i > 1$. For that agent~$i$, the
    maximum term in~\cref{eq:analysis:bbr1:queue-condition} is exactly 1, and
    the maximum term in~\cref{eq:analysis:bbr1:queue-condition} is smaller than 1,
    which makes the conditions contradictory and rules out an equilibrium.
    Conversely, if assuming an equilibrium 
    where $\exists i\in U_{\mathrm{\ell^{\ast}}}.\ \Delta_i < 1$,
    the maximum term in~\cref{eq:analysis:bbr1:queue-condition} is~$1/\Delta_i > 1$, 
    and the maximum term in~\cref{eq:analysis:bbr1:rate-condition} is also~$1/\Delta_i$,
    which again leads to contradictory equations.
    Hence, no equilibria other than the equilibrium 
    with~$\forall i\in U_{\mathrm{\ell^{\ast}}}.\ \Delta_i = 1$ are possible,
    which concludes the proof.

\subsection{Proof of~\cref{thm:analysis:bbr1:stability}}
\label{sec:proof:bbr1:stability}

In the scenario under consideration, the equilibrium requires that
    the propagation delay~$d_i$ is equal for all senders.
    Moreover, it holds that~$q_{\ell} = 0\ \forall \ell \neq \ell^{\ast}$.
    Hence, we can simplify: 
    $\Delta_i = \Delta(q_{\ell^{\ast}}) := 2d / (d+q_{\ell^{\ast}}/C_{\ell^{\ast}})$,
    where~$d$ is the propagation delay experienced by all agents. As a result,
    the equilibrium requires that~$\Delta(q_{\ell^{\ast}}) = 1 \iff q_{\ell^{\ast}} = dC_{\ell^{\ast}}$.
    
    We now consider a configuration where the senders are out of equilibrium and constrained
    by the congestion-window limit, i.e., $\Delta(q_{\ell^{\ast}}) < 1 \iff q_{\ell^{\ast}} > dC_{\ell^{\ast}}$.
    The dynamics of the bottleneck-bandwidth estimates by sender~$i$ are then given by:
    \begin{equation}
        \dot{x}_i^{\mathrm{btl}} = \frac{\Delta(q_{\ell^\ast})x_i^{\mathrm{btl}}C_{\ell^\ast}}{\Delta(q_{\ell^\ast})\sum_{k\in U_{\ell^\ast}} x_k^{\mathrm{btl}}} - x_i^{\mathrm{btl}}
        = \frac{x_i^{\mathrm{btl}}C_{\ell^\ast}}{\sum_{k\in U_{\ell^\ast}} x_k^{\mathrm{btl}}} - x_i^{\mathrm{btl}}
    \end{equation} Moreover, the dynamics of the bottleneck-link queue~$q_{\ell^\ast}$ are:
    \begin{equation}
        \dot{q}_{\ell^\ast} = y_{\ell^\ast} - C_{\ell^\ast} = \Delta(q_{\ell^\ast})\sum_{i\in U_{\ell^\ast}} x_i^{\mathrm{btl}} - C_{\ell^\ast}
    \end{equation} where~$y_{\ell^\ast}$
    is the arrival rate at bottleneck link~$\ell^\ast$.
    Based on these dynamics, we derive the dynamics of the 
    arrival rate~$y_{\ell^\ast}$:
    \begin{align}
        \dot{y}_{\ell^\ast} &= \dot{\Delta}(q_{\ell^\ast})\sum_{i\in U_{\ell^\ast}} x_i^{\mathrm{btl}} + \Delta(q_{\ell^\ast})\sum_{i\in U_{\ell^\ast}} \dot{x}_i^{\mathrm{btl}}\\
        % &= - \frac{2d\dot{q}_{\ell^\ast}}{C_{\ell^\ast}(d + \frac{q_{\ell^\ast}}{C_{\ell^\ast}})^2} \sum_{i\in U_{\ell^\ast}} x_i^{\mathrm{btl}} + \Delta(q_{\ell^\ast}) \sum_{i\in U_{\ell^\ast}} \frac{x_i^{\mathrm{btl}}C_{\ell^\ast}}{\sum_{k\in U_{\ell^\ast}} x_k^{\mathrm{btl}}} - \Delta(q_{\ell^\ast}) \sum_{i\in U_{\ell^\ast}} x_i^{\mathrm{btl}}\\
        % &= -\frac{y_{\ell^\ast}-C_{\ell^\ast}}{C_{\ell^\ast}(d + \frac{q_{\ell^\ast}}{C_{\ell^\ast}})} y_{\ell^\ast} + \Delta(q_{\ell^\ast}) C_{\ell^\ast} - y_{\ell^\ast} 
        &= -\frac{1}{C_{\ell^\ast}(d + \frac{q_{\ell^\ast}}{C_{\ell^\ast}})} y_{\ell^\ast}^2 + (\frac{1}{d + \frac{q_{\ell^\ast}}{C_{\ell^\ast}}} - 1)y_{\ell^\ast} + \Delta(q_{\ell^\ast})C_{\ell^\ast}\nonumber
    \end{align}
    
    Building on this formalization, we can define a classic non-linear dynamic system  
    with~$y_{\ell^\ast}$ and~$q_{\ell^\ast}$ as state variables, and
    ~$\dot{y}_{\ell^\ast}$ and~$\dot{q}_{\ell^\ast}$ as entries of the vector-valued function~$f$
    describing the dynamics. To characterize the stability of that system, we can then employ 
    the indirect Lyapunov method~\cite{pukdeboon2011review}. This method states that
    a system is locally asymptotically stable if the Jacobian matrix of the system dynamics~$f$
    has eigenvalues with exclusively negative real parts when evaluated at the equilibrium.
    The Jacobian matrix~$\mathbf{J}_f \in \mathbb{R}^{2\times2}$ has 
    the following entries:
    \begin{equation}
        \begin{split}
            \frac{\partial\dot{y}_{\ell^\ast}}{\partial y_{\ell^\ast}} 
        &= -\frac{2}{C_{\ell^\ast}(d + \frac{q_{\ell^\ast}}{C_{\ell^\ast}})} y_{\ell^\ast} + \frac{1}{d + \frac{q_{\ell^\ast}}{C_{\ell^\ast}}} - 1 \\
        \frac{\partial\dot{y}_{\ell^\ast}}{\partial q_{\ell^\ast}} 
        &= \frac{y_{\ell^\ast}^2}{C_{\ell^\ast}^2(d+\frac{q_{\ell^\ast}}{C_{\ell^\ast}})^2} - \frac{y_{\ell^\ast}}{C_{\ell^\ast}(d+\frac{q_{\ell^\ast}}{C_{\ell^\ast}})^2} - 
        \frac{2d}{(d+\frac{q_{\ell^\ast}}{C_{\ell^\ast}})^2}\\
        \frac{\partial\dot{q}_{\ell^\ast}}{\partial y_{\ell^\ast}} &= 1  \quad \frac{\partial\dot{q}_{\ell^\ast}}{\partial q_{\ell^\ast}} = 0
        \end{split}
    \end{equation}
    
    Evaluating this Jacobian matrix at the equilibrium, i.e., $y_{\ell^\ast} = C_{\ell^\ast}$ and~$q_{\ell^\ast} = dC_{\ell^\ast}$,
    yields a matrix for which the maximum eigenvalue~$\lambda^+$ can be found via the characteristic equation:
    \begin{equation}
    \begin{split}
        &\mathbf{J}_f(C_{\ell^\ast},\ dC_{\ell^\ast}) = \begin{pmatrix} 
        -\frac{1}{2d} - 1 & -\frac{1}{2d} \\
        1 & 0
        \end{pmatrix}\\
        \implies &\quad 
        \lambda^+ = -(\frac{1}{4d}+\frac{1}{2}) + \frac{1}{2d} ( \pm(d-\frac{1}{2}) ) 
    \end{split}
    \end{equation}
    
    Performing a case distinction on~$d$ confirms that the maximum eigenvalue~$\lambda^+$
    is always negative:
    \begin{equation}
        \begin{split}
            d \leq \frac{1}{2}: \quad &\lambda^+ = \frac{ -(\frac{1}{2d}+1) - \frac{1}{d} (d-\frac{1}{2}) }{ 2 } = -1 < 0\\
            d > \frac{1}{2}: \quad &\lambda^+ = \frac{ -(\frac{1}{2d}+1) + \frac{1}{d} (d-\frac{1}{2}) }{ 2 } = -\frac{1}{2d} < 0\\
        \end{split}
    \end{equation}
    Hence, we observe that the Jacobian matrix of the system dynamics has consistently negative eigenvalues, 
    which by the indirect Lyapunov method proves the asymptotic stability of the dynamics.

\subsection{Proof of~\cref{thm:analysis:bbr1:equilibrium:shallow}}

    Given~$\Delta_i \geq \nicefrac{5}{4}$ for all~$i \in U_{\ell^\ast}$,
    the equilibrium condition on~$\{x_i^{\mathrm{btl}}\}_{i\in U_{\ell^\ast}}$
    is:
    \begin{equation}
        \forall i \in U_{\ell^\ast}.\ x_i^{\mathrm{btl}} = \frac{\nicefrac{5}{4}x_i^{\mathrm{btl}}C_{\ell^\ast}}{\nicefrac{5}{4}x_i^{\mathrm{btl}} + \sum_{j\neq i} x_j^{\mathrm{btl}}} = C - \nicefrac{4}{5} \sum_{j\neq i} x_j^{\mathrm{btl}} 
    \end{equation}
    This equation system requires all~$\{x_i^{\mathrm{btl}}\}_{i\in U_{\ell^\ast}}$
    to be equal, which allows a straightforward solution:
    \begin{equation}
        \forall i \in U_{\ell^\ast}.\ x_i^{\mathrm{btl}} =  \frac{\nicefrac{5}{4}x_i^{\mathrm{btl}}C_{\ell^\ast}}{(N+\nicefrac{1}{4})x_i^{\mathrm{btl}}} = \frac{5C_{\ell^\ast}}{4N+1}
    \end{equation}
    
    It remains to show that this equilibrium is asymptotically stable,
    for which we employ the indirect Lyapunov method.
    We apply this method to a non-linear dynamic process
    with~$\{x_i^{\mathrm{btl}}\}_{i\in U_{\ell^\ast}}$ as state variables
    and~$\{\dot{x}_i^{\mathrm{btl}}\}_{i\in U_{\ell^\ast}}$ as vector-valued
    evolution function~$f$. The Jacobian matrix~$\mathbf{J}_f$
    has the following entries, which we evaluate at the equilibrium:
    \begin{align}
        \frac{\partial\dot{x}_i}{\partial x_i} &= 
        \frac{\nicefrac{5}{4} C\sum_{j\neq i} x_j^{\mathrm{btl}}}{(\nicefrac{5}{4}x_i+\sum_{j\neq i} x_j)^2} - 1
        \stackrel{Eq.}{=} -\frac{5}{4N+1} =: J_{ii}\\
        \frac{\partial\dot{x}_i}{\partial x_j} &= 
        - \frac{\nicefrac{5}{4} Cx_i}{(\nicefrac{5}{4}x_i+\sum_{j\neq i} x_j)^2}
        \stackrel{Eq.}{=} -\frac{4}{4N+1} =: J_{ij}\\
    \end{align}
    The eigenpairs~$(\lambda, \mathbf{v})$ of $\mathbf{J}_f$ at the equilibrium
    satisfy the following conditions:
    \begin{equation}
        \forall i\in U_{\ell^\ast}.\ (J_{ii}-\lambda)v_i + J_{ij}\sum_{j\neq i} v_j = 0
    \end{equation}
    The first type of solution for this equation system is given by $\lambda = J_{ii} - J_{ij} < 0$ and 
    every~$\mathbf{v}$ with~$\lVert \mathbf{v} \rVert_1 = 0$. The second type of solution is found
    by assuming~$\lambda \neq J_{ii} - J_{ij}$, which implies equal~$v_i$ $\forall i \in U_{\ell^\ast}$
    and hence (together with~$v_i \neq 0$) $\lambda = J_{ii} + (N-1)J_{ij} < 0$.
    Since the eigenvalues of the Jacobian are thus consistently negative,
    the indirect Lyapunov method suggests that the dynamics are asymptotically stable.

\subsection{Proof of~\cref{thm:analysis:bbr2:equilibrium}}

The equilibrium conditions  can
    be translated into the following constraints
    given~$q_{\ell^\ast} > 0$:
    \begin{align}
         x_i^{\mathrm{btl}} &= \max(1,\ \nicefrac{1}{\delta_i}) \cdot (C_{\ell^{\ast}} - \sum_{j\neq i} \min(1,\ \delta_j) \cdot x_j^{\mathrm{btl}}) \label{eq:analysis:bbr2:queue-condition}\\
        x_i^{\mathrm{btl}} &= C_{\ell^{\ast}} - \nicefrac{4}{5} \cdot \max(1,\ \nicefrac{1}{\delta_i}) \cdot \sum_{j\neq i} \min(1,\ \delta_j) \cdot x_j^{\mathrm{btl}} \label{eq:analysis:bbr2:rate-condition}
    \end{align}
    While these constraints potentially admit multiple equilibria, the equilibrium 
    from~\cref{thm:analysis:bbr2:equilibrium}
    is a special equilibrium for which~$\delta_i$ is equal across all~$i\in U_{\ell^\ast}$, i.e., $\delta_i = \delta$.
    Substituting~$\delta$ for all~$\delta_i$, and 
    equating~\cref{eq:analysis:bbr2:queue-condition} with~\cref{eq:analysis:bbr2:rate-condition},
    which first yields:
    \begin{equation}
        \begin{split}
        &\forall i\in U_{\ell^\ast}.\ 
        \sum_{j\neq i} x_j^{\mathrm{btl}} = 5\cdot \left(\max(1,\ \nicefrac{1}{\delta}) - 1\right) \cdot C \\
        \implies \quad 
        &\forall i \in U_{\ell^\ast}.\ x_i^{\mathrm{btl}} = \max(1,\ \nicefrac{1}{\delta})\cdot\frac{C}{N},
        \label{eq:analysis:bbr2:equal-share}
        \end{split}
    \end{equation} where the equation systems
    requires that all~$x_i^{\mathrm{btl}}\ \forall i\in U_{\ell^\ast}$ equal the same value
    (hence perfect fairness), 
    where this value can be found 
    using~\cref{eq:analysis:bbr2:queue-condition}.
    By inserting~$x_i^{\mathrm{btl}}$
    from~\cref{eq:analysis:bbr2:equal-share}
    into~\cref{eq:analysis:bbr2:rate-condition}, 
    it can be shown that $\delta \leq 1$ by
    producing a contradiction for~$\delta > 1$.
    In contrast, solving that equation given~$\delta \leq 1$ yields $\delta = \frac{4N+1}{5N}$,
    which is equivalent to the condition in~\cref{thm:analysis:bbr2:equilibrium}.
    This insight concludes the proof.

\subsection{Proof of~\cref{thm:analysis:bbr2:stability}}

 In the scenario under consideration, the equilibrium requires that
    the propagation delay~$d_i$ is equal for all senders.
    Moreover, it holds that~$q_{\ell} = 0\ \forall \ell \neq \ell^{\ast}$.
    Hence, we can simplify: 
    $\delta_i = \delta(q_{\ell^{\ast}}) := d / (d+q_{\ell^{\ast}}/C_{\ell^{\ast}})$,
    where~$d$ is the propagation delay experienced by all agents. As a result,
    the equilibrium requires that~$\delta(q_{\ell^{\ast}}) = \frac{4N+1}{5N} \iff q_{\ell^{\ast}} = \frac{N-1}{4N+1}dC_{\ell^{\ast}}$.
    
    We translate the reduced model from~\cref{sec:analysis:bbr2:reduction} into
    a nonlinear dynamic process with the sending
    rates~$\{x_i\}_{i\in U_{\ell^\ast}}$ and the queue length~$q_{\ell^\ast}$ as state variables.
    The evolution of these state variables
    is given by vector-valued function~$f$
    with the following entries:
    \begin{align}
        \dot{x}_i &= \dot{\delta}(q_{\ell^\ast})x_i^{\mathrm{btl}} + \delta(q_{\ell^\ast})\dot{x}_i^{\mathrm{btl}} \\
        &= \left( \frac{C_{\ell^\ast}-\sum_{k\in U_{\ell^\ast}} x_k}{C_{\ell^\ast}(d+q_{\ell^\ast}/C_{\ell^\ast})} + \frac{\nicefrac{5}{4}\delta C}{\nicefrac{5}{4}x_i+\sum_{j\neq i} x_j} - 1\right)\cdot x_i \nonumber\\
        \dot{q}_{\ell^\ast} &= \sum_{i\in U_{\ell^\ast}} x_i - C_{\ell^\ast}
    \end{align}
    
    The corresponding Jacobian matrix~$\mathbf{J}_f$
    is composed of the following entries:
    \begin{align}
        \frac{\partial\dot{x}_i}{\partial x_i} &= 
        \frac{C_{\ell^\ast}-2x_i-\sum_{j\neq i} x_j}{C_{\ell^\ast}(d+q_{\ell^\ast}/C_{\ell^\ast})} + \frac{\nicefrac{5}{4}\delta C\sum_{j\neq i} x_j^{\mathrm{btl}}}{(\nicefrac{5}{4}x_i+\sum_{j\neq i} x_j)^2} - 1  \\
        \frac{\partial\dot{x}_i}{\partial x_j} &= 
        -\frac{x_i}{C_{\ell^\ast}(d+q_{\ell^\ast}/C_{\ell^\ast})} - \frac{\nicefrac{5}{4}\delta Cx_i}{(\nicefrac{5}{4}x_i+\sum_{j\neq i} x_j)^2} \\
        \frac{\partial\dot{x}_i}{\partial q} &= \frac{1}{d+\frac{q_{\ell^\ast}}{C_{\ell^\ast}}} \left( \frac{C_{\ell^\ast}-\sum_{k\in U_{\ell^\ast}} x_k}{C_{\ell^\ast}^2\left(d+\frac{q^{\ell^\ast}}{C_{\ell^\ast}}\right)} - \frac{\nicefrac{5}{4} \delta C}{\nicefrac{5}{4}x_i+\sum_{j\neq i} x_j} \right) x_i\\
        \frac{\partial\dot{q}}{\partial x_i} &= 1 \quad \frac{\partial\dot{q}}{\partial q} = 0
    \end{align}
    
    Evaluating the Jacobian matrix at
    the equilibrium point from~\cref{thm:analysis:bbr2:equilibrium}
    yields the following matrix~$\mathbf{J}$:
    \begin{align}
         \frac{\partial\dot{x}_i}{\partial x_i} &=
         -\frac{4N+1}{5N^2d} - \frac{5}{4N+1} =: J_{ii} \quad 
         & \frac{\partial\dot{q}}{\partial x_i} &= 1\\
         \frac{\partial\dot{x}_i}{\partial x_i} &=
         -\frac{4N+1}{5N^2d} - \frac{4}{4N+1} =: J_{ij} \quad 
         & \frac{\partial\dot{q}}{\partial q} &= 0\\
         \frac{\partial\dot{x}_i}{\partial q} &=
         -\frac{4N+1}{5N^2d} =: J_{iq}
    \end{align}
    
    By Lyapunov's indirect method, the above
    Jacobian matrix must have exclusively
    negative eigenvalues in order for the
    equilibrium to be asymptotically stable, i.e.,
    for every pair~$(\lambda,\mathbf{v})$ with~$\mathbf{Jv} = \lambda\mathbf{v}$,
    the eigenvalue $\lambda$ must be lower 
    than~0. To verify this property~$\mathbf{J}$,
    we concretize the eigenvalue condition:
    \begin{align}
        &\forall i \in U_{\ell^\ast}. \quad J_{ii} v_i + J_{ij} \sum_{j\neq i} v_j + J_{iq} v_q = \lambda v_i\\
        &\sum_{i \in U_{\ell^\ast}} v_i = \lambda v_q
    \end{align}
    
    By solving these equations for $v_q$ and equating
    the resulting terms, we obtain the following
    conditions:
    \begin{equation}
        \forall i \in U_{\ell^\ast}.\ \sum_{k\in U_{\ell^\ast}} v_k = \frac{\lambda}{J_{iq}}\left( (\lambda - J_{ii})v_i - J_{ij}\sum_{j\neq i}v_j \right)
        \label{eq:analysis:bbr2:eigen-system}
    \end{equation} This equation allows two types
    of solutions.
    First, for~$\lambda = J_{ii} - J_{ij}$,
    the set of valid eigenvectors~$\mathbf{v}$ 
    is only constrained by a condition on~$\sum_{k \in U_{\ell^\ast}} v_k$; more importantly for
    the proof, $\lambda$ is negative.
    Second, for~$\lambda \neq J_{ii} - J_{ij}$,
    the values~$v_i\ \forall i\in U_{\ell^\ast}$
    must be equal such that the
    equation system from~\cref{eq:analysis:bbr2:eigen-system} can be collapsed into
    a single quadratic equation, which yields the maximum eigenvalue~$\lambda^+$:
    \begin{equation}
        N\cdot v_i = \frac{\lambda}{J_{iq}}\left( (\lambda - J_{ii})v_i - J_{ij}(N-1)v_i \right) \implies \lambda^+ = -1
    \end{equation}
    Since the maximum eigenvalue~$\lambda^+$ is
    negative, all eigenvalues of~$\mathbf{J}$ are
    negative, which by the indirect Lyapunov
    method proves that the dynamic process
    defined by~$f$ is asymptotically stable.

\end{document}